\documentclass[pre,twocolumn,showpacs,preprintnumbers,amsmath,amssy mb]{revtex4}

\usepackage{graphics}

\usepackage{graphicx}
\usepackage{dcolumn}
\usepackage{bm}

\newcommand{\beq}{\begin{equation}}
\newcommand{\eeq}{\end{equation}}    

\newcommand{\bea}{\begin{eqnarray}}
\newcommand{\eea}{\end{eqnarray}}    

\begin{document}

\font\small=cmr8           
\font\petit=cmcsc10        

\font\bbf=cmbx10 scaled\magstep1 
\font\bbbf=cmbx10 scaled\magstep2 
\font\bbbbf=cmbx10 scaled\magstep3 

\def\subti#1{\par\vskip0.8cm{\bf\noindent #1}\par\vskip0.4cm}
\def\ti#1{\par\vskip1.6cm{\bbf\noindent #1}\par\vskip0.8cm}
\def\bigti#1{\par\vfil\eject{\bbbf\noindent #1}\par\vskip1.6cm}

\def\cbigti#1{\par\vskip2.0cm{\bbbf\noindent\centerline {#1}\par\vskip0.5cm}}
\def\cti#1{\par\vskip1.0cm{\bbf\noindent\centerline {#1}\par\vskip0.4cm}}
\def\csubti#1{\par\vskip0.5cm{\bf\noindent\centerline {#1}\par\vskip0.3cm}}

\def\doublespace {\baselineskip 22pt}        

\def\eqd{\buildrel \rm d \over =}    
\newcommand{\p}{\partial}           
\def\px{\partial _x}           
\def\py{\partial _y}           
\def\pz{\partial _z}           
\def\pt{\partial _t}           
\def\ssum{\textstyle\sum}
\def\arr{\rightarrow}
\def\id{\equiv}
\def\eqv{\leftrightarrow}
\def\fol{\rightarrow}
\let\prop=\sim
\def\gapprox{\;\rlap{\lower 2.5pt            
 \hbox{$\sim$}}\raise 1.5pt\hbox{$>$}\;}       
\def\lapprox{\;\rlap{\lower 2.5pt            
 \hbox{$\sim$}}\raise 1.5pt\hbox{$<$}\;} 

\def\ang{\,{\rm\AA}}
\def\cm{\,{\rm cm}}
\def\km{\,{\rm km}}
\def\kpc{\,{\rm kpc}}
\def\second{\,{\rm sec}}     
\def\erg{\,{\rm erg}}
\def\ev{\,{\rm e\kern-.1em V}}
\def\kev{\,{\rm ke\kern-.1em V}}
\def\k{\,{\rm K}}
\def\K{\,{\rm K}}
\def\gauss{\,{\rm gauss}}
\def\SFU{\,{\rm SFU}}

\newcommand{\R}{\,{\rm I\kern-.15em R}}
\def\N{\,{\rm I\kern-.15em N}}
\def\N{\,{\rm /\kern-.15em R}}

\def\mhz{\,{\rm MHz}}

\def\n{\noindent}
\def\lead{\leaders\hbox to 10pt{\hfill.\hfill}\hfill}
\def\a{\"a}
\def\o{\"o}
\def\u{\"u}
\def\infinit{\infty}
\def\upr#1{\rm#1}

\def\dd{D^{\left(2\right)}}
\def\cc{C_d ^{\left(2\right)} \left(\epsilon\right)}


\newcount\glno
\def\no{\global\advance\glno by 1 \the\glno}

\newcount\secno
\def\newsec{\global\advance\secno by 1}
\def\sec{\the\secno}

\newcount\chapno
\def\newchap{\global\advance\chapno by 1}
\def\chap{\the\chapno}



\preprint{....}

\title{Random walk through fractal environments}

\author{H.\ Isliker}
\email{isliker@helios.astro.auth.gr}
\author{L.\ Vlahos}
\email{vlahos@helios.astro.auth.gr}

\affiliation{
Association Euratom--Hellenic Republic \\
Section of Astrophysics, Astronomy and Mechanics \\
Department of Physics, University of Thessaloniki \\
GR 54006 Thessaloniki, GREECE }

\date{\today}

\date{Received ...; accepted ...}


\begin{abstract}
We analyze random walk through fractal environments, 
embedded in 3--dimensional, permeable
space. Particles travel freely and are scattered off into random directions
when they hit the fractal.
The statistical 
distribution of the flight increments (i.e.\ of the displacements between 
two consecutive hittings) is 
analytically derived from a common, practical definition of fractal dimension, 
and it turns out to approximate quite well a power-law
in the case where the dimension $D_F$ of the fractal 
is less than 2, there is though always a finite rate of unaffected escape. 
Random walks through fractal sets with $D_F\le 2$ 
can thus be considered as defective Levy walks. 
The distribution of jump increments for $D_F > 2$ is decaying exponentially.
The diffusive behavior 
of the random walk is analyzed in the frame of continuous time random walk, 
which we generalize to include the case of defective distributions of 
walk-increments.
It is shown that the particles 
undergo anomalous, enhanced diffusion for $D_F < 2$,
the diffusion is dominated by the finite escape rate.
Diffusion for $D_F>2$ is normal for large times, enhanced though 
for small and intermediate times.
In particular, it follows that fractals generated by a particular class 
of self-organized 
criticality (SOC) models give rise to enhanced diffusion.
The analytical results are illustrated by Monte-Carlo simulations.
\end{abstract}



\maketitle

\section{Introduction}

We study the problem of particles performing a random walk through
a fractal environment in 3-dimensional embedding space. 
The particles travel freely in the space not occupied by the fractal and
are scattered off into random directions when they hit the fractal.
We derive analytically the distribution $p_r$ of the random walk 
increments as a function of the dimension $D_F$ of the fractal set, and 
we calculate the diffusivity analytically, using
the formalism of continuous time random walk (CTRW; e.g.\ 
\cite{Montroll1965}), which we generalize here in order to include
the case of defective (not normalized to one) distributions of 
walk-increments.
The random walk is finally illustrated by Monte Carlo simulations.


The physical applications for the theory developed here are 
to systems consisting of a large number of spatially distributed, 
localized scatterers (accelerators), whose support forms 
a fractal set, suspended in a permeable medium, and in which  
particles move, with their dynamics being governed by collisions 
with the fractal:
The particles move freely in the system except when they hit a
part of the fractal (a scatterer), where they undergo the respective 
interaction, after which
they leave the scattering center, possibly hit the fractal again,
and so forth, performing thus a random walk in between 
subsequent interactions. 

One application of the introduced theory is to particle transport in 
turbulent plasmas, whenever it can be asserted that the field inhomogeneities 
are distributed in a fractal way. This is implicitly claimed there
where turbulent plasmas have successfully been modeled 
with self-organized criticality (SOC; about SOC see \cite{BTW}):
In Ref.\ \cite{Isliker2001}, it has recently been shown that the unstable 
sites at temporal snap-shots during an avalanche in 3D form a  
fractal with dimension roughly $1.8$. 
This can be expected to hold for all SOC models whose evolution 
rules are of the type of \cite{Lu}.
Particles moving in the model will thus undergo the type of diffusion
we analyze here.

Concrete examples of applications include the following:
(i) Solar flares have been shown to be compatible with SOC (\cite{Lu,Isliker}).
The unstable sites of the SOC model,
which represent small-scale current-dissipation regions (see \cite{Isliker}), 
cause the acceleration of particles, which perform thus a random walk 
of the type we analyze here. 
(ii) Though the question is still under debate, there are indications
that the Earth's magnetosphere exhibits structures
compatible with SOC (e.g.\ \cite{magentosphere}).
(iii) In inquiries on confined plasmas and the related transport phenomena, 
evidence has been collected that the confined plasma might be in the state 
of SOC (claimed in \cite{SOC}, doubted though in \cite{Carbone}). 
Moreover, it is known that particles in confined plasmas
undergo anomalous diffusion (\cite{confined}), 
a property which we will show also to hold often for 
the particles in the kind of systems we analyze here.
(iv) A non-plasma, non-SOC example, where 
the theory developed here potentially can be applied, is
the random walk of cosmic particles, which are scattered off the 
fractally distributed galaxies (e.g.\ \cite{Bak2001}).

The investigation we present here is to be contrasted to two 
related, though characteristically different kinds of studies:
--- (I) In Ref.\ \cite{frac}, random walks and diffusion 
\textit{along} fractals are investigated, where the particles are forced to 
move along a fractal structure.
The fractals investigated are connected fractals or percolation backbones.
These studies are motivated by applications to the transport in porous media, 
or along percolation networks, and it is established that the diffusion in 
these systems is often anomalous.
Differently to these studies, our random walkers 
cross the permeable space freely, they are not forced to follow the 
fractal structure, but they just occasionally hit the fractal. 
--- (II) In Ref.\ (\cite{sand}), the random walk 
of sand grains in sand-pile (SOC) models is investigated, i.e.\ the direct 
transport of the unstable sites, which are found to undergo anomalous,
enhanced diffusion. In contrast to these studies, when applying 
our theory to 
SOC models, we do not study the diffusion of the unstable sites
(the transport of sand grains),
but the diffusion of additional particles, foreign to the system 
(they are not contained in pure sand-pile models), which interact with the 
unstable sites, i.e.\ we freeze time in the
avalanche model and let particles interact with the spatially distributed
unstable sites. This is motivated through applications 
where the sand-pile does not model the evolution of real sand or rice-piles, 
but where it models ultimately the evolution of some kind of forces
in dilute media (e.g.\ some kind of stress forces, or the magnetic 
or electric field in plasmas), in which the avalanche model merely gives
the locations of the instabilities which affect particles moving otherwise
freely in the system. 

The fractal sets which constitute the environment we analyze
are {\it natural} fractals, which exhibit self-similar
scaling behavior only in a finite range, and which are made up
of finite, 3-dimensional elementary volumes,
small in size compared to the size of the fractal. 
According to Ref.\ \cite{Sinai1980}, such environments could be termed 
3-dimensional, fractal Lorentz gas.
Actually, any fractal set encountered in nature is 
a natural fractal in the sense introduced here, from the classical 
examples (the coast-line
of Britain, cloud-surfaces, etc.; see \cite{Mandelbrot1982}),
to the localized scattering centers of the above mentioned 
applications mainly to plasma physics, which are yet small regions, 
with finite volumes, though. 

In Sec.\ \ref{SecII}, we will specify the notion of natural fractals, 
introduce the way we model the fractal scaling behaviour of natural fractals, 
derive analytically the probability distribution of the random walk 
increments for random walks through fractal environments, 
give the relations for the rate of unaffected escape from the system, and 
derive
approximate forms of the distribution of jump increments.
In Sec.\ \ref{SecIII}, the theory of Continuous Time Random Walk 
(CTRW) will be introduced and generalized to include the case of defective 
(not normalized to one) distributions of jump increments. 
The CTRW formalism will then be applied to determine analytically the 
diffusive 
behaviour for random walks through fractal environments,
and to calculate the expected number of collisions with the fractal 
(in Sec.\ \ref{SecIV} for fractal dimensions $D_F<2$, and in Sec.\ 
\ref{SecV} for $D_F>2$).
In Sec.\ \ref{SecVI}, the analytical 
results will be compared to and illustrated by Monte-Carlo simulations.
The results are summarized and discussed in Sec.\ \ref{SecVII},
and Conclusions are drawn in Sec.\ \ref{SecVIII}.

\section{Probability distribution of the increments of a random walk
through a fractal environment \label{SecII}}

\subsection{Specification of the problem; natural fractals\label{SecIIA}}

\begin{figure}
\resizebox{\hsize}{!}{\includegraphics{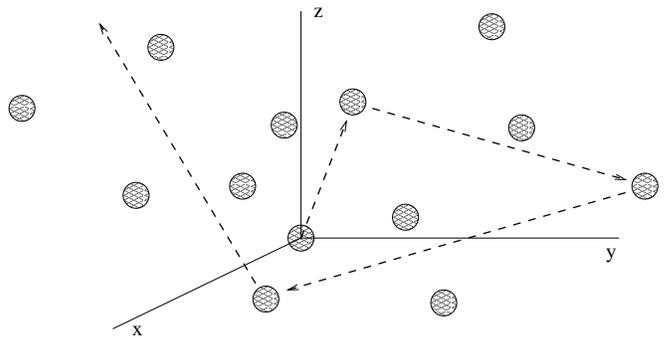}}
\caption{Sketch: The random walk in 3-dimensional space we analyze.
A particle trajectory is indicated with an arrow. 
The small shaded regions are the elementary volumes $\delta V$
the fractal consists of (see Sec.\ \ref{SecIIA}).
\label{fig0} }
\end{figure}

We assume a fractal $F$ embedded in 3-dimensional space ($\R^3$)
with a fractal dimension $D_F$ (e.g.\ box-counting or correlation 
dimension). We furthermore assume that there are particles (the random
walkers) which travel freely in the empty 
(in the sense of not affecting the random walkers) space, but are scattered 
into random directions off the points belonging to the fractal $F$.
Fig.\ \ref{fig0} sketches the situation.
Our interest in this Section is in the statistical distribution of the 
particles' traveled distances in between two consecutive collisions with 
the fractal, i.e.\ of the random walk increments.

If we pose the problem in this form, then the fractals fall into 
two distinct classes: 
Imagine a particle to be situated at a point $\vec x_i$ belonging 
to $F$, somewhere in the interior. The particle actually 'sees'
the projection of the fractal onto a large imaginary sphere $S$ around
$F$ and centered at $\vec x_i$, exactly as we do see the stars projected 
onto the celestial sphere. This sphere is 2-dimensional, so that, 
if $D_F<2$, the projection $F_P$ of $F$ onto $S$ has dimension
$D_P=D_F<2$ (see e.g.\ \cite{Falconer1990}). This implies that $F_P$ has 
zero measure (no volume). The possible trajectories for the particle 
are the straight lines originating from $\vec x_i$. 
The probability of such a trajectory to hit the fractal
$F$ at all is the area occupied by $F_P$ on $S$, divided by the area of $S$;
thus, the probability to hit the fractal is zero, the particle will almost
never hit the fractal, it will almost surely escape from the system,
and it does not make sense to determine a distribution of random walk 
increments.

On the other hand, if $D_F\geq 2$, then the projection of $F$ onto
$S$ has dimension $D_P=2$, and the area occupied by $F_P$ on $S$
is positive (see e.g.\ \cite{Falconer1990}). The probability of the particle to
hit the fractal, which
is again the area of $F_P$ on $S$, divided by the area of $S$, is finite,
and it makes sense to determine a distribution of walk increments
(for an isotropic fractal, we expect the probability to hit the fractal
to be $1$, but
there may be a finite probability for a particle to escape
unaffected, without hitting the fractal at all, depending on the degree 
of spatial
anisotropy of the concrete fractal under consideration). 
 
This distinction holds for mathematical fractals, which per definition
exhibit a scale-free scaling behavior (self-similarity or --- 
possibly statistical --- self-affinity) from their usually finite size
down to all scales. Our interest here is though in what we term 
{\it natural} fractals: They are characterized by the following 
properties:
(i) They are sampled only with a finite number of points.
(ii) Their scaling behavior exhibits a lower cut-off, 
i.e.\ irrespective of the numerical method used to determine
their fractal dimension, there will be a lower limit of scaling 
for the estimator. Correspondingly, there is a finite minimum separation
distance $\delta$ between the points of the fractal. 
This property is partly a consequence of property (i).
(iii) The elements the natural fractals consists of are not mathematical
points, line-segments, or surface elements,
but they represent finite, yet small 3-dimensional elementary volumes 
$\delta V$, at most of radial size $\delta/2$.

Properties (i) and (ii) characterize what one might call a {\it finite}
fractal. They imply that $F=\{x_i\}_{i=1,...,n_F}$, i.e.\ $F$ is a finite 
collection of $n_F$ isolated points, with $\delta$ the smallest distance 
between them. If we assume the set $F$ to be contained in a sphere with
radius $l$, then $F$ exhibits a fractal scaling behaviour for scales $r$ in
the range $\delta\leq r\leq l$. The property (iii) makes the fractal natural
in the sense that the points $x_i$ of $F$ represent actually small 
3-dimensional volumes 
$\delta V$ of radial sizes smaller than $\delta/2$, with which a particle can
interact through some forces, depending on the concrete physical application. 
We assume correspondingly an interaction cross-section
$\rho^2\pi$ with cross-sectional radius $\rho$
to be associated with every point belonging to the fractal.

Since the $\delta V$ are 3-dimensional objects, we must require that
that the volumes $\delta V$ should be smaller in radial size ($\rho$) than 
$\delta/2$, i.e.\ $\rho \leq \delta/2$:
If $\rho$ were larger than
$\delta/2$, then the fractal scaling of the natural fractal 
would break down already at the scale $2\rho$, the diameter of the 
elementary volumes $\delta V$, 
i.e.\ before reaching the scale $\delta$, which means that $\delta$ would
have been inadequately determined and would have to be adjusted.
Moreover, if the radius $\rho$ of the volumes $\delta V$ were in the range
$\delta/2 \leq \rho \leq \delta$, then near elementary volumes would 
over-lap, and they would be taken 
for one elementary volume.

In the frame of natural fractals, the random walk problem we pose takes
a different shape: if the fractal were just finite, then all the particles
would almost surely escape from the system without colliding with the fractal,
since the probability to hit a finite set of isolated points with a straight
line trajectory is obviously zero (the finite fractal in any case is a set 
of measure (volume) zero). Yet, since the fractals we analyze
are natural, the isolated points of the fractal represent finite 
3-dimensional volumes with a corresponding finte cross section,
so that there is a finite probability for a particle to collide with these 
elementary volumes, and it makes sense to determine the corresponding 
distribution of walk increments.

A clarification is to be made concerning the scattering process:
The scattering of the particles off the points of the fractal 
(the elementary volumes) is not scattering off hard spheres.
We consider the elementary volumes as regions into 
which particles can penetrate, they will though be affected 
by some forces inside these regions. This is realistic since 
our main application is to plasma-physics, where the elementary 
volumes are typically regions where an electric field resides.

Last, we note that the radial size $l$ of the entire fractal 
is of course finite in any reasonable physical application.
The derivations we will give in the following are consequently 
made for the case of finite fractals ($l<\infty$), it will though
turn out that $l$ appears just as an arbitrary exterior parameter  
and therewith is allowed to take arbitrarily large values
(Secs.\ \ref{SecIIC1} and \ref{SecIIC2}).
It will furthermore turn out that several characteristics
of the problem we analyze
assume finite asymptotic values if $l$ becomes very large 
(Secs.\ \ref{SecIID} and \ref{SecIIE}).
We will thus include in our treatment the case of what we term 
{\it asymptotically large} systems, by which we mean systems where
$l$ is so large that the asymptotic, large $l$ behaviour is
practically reached, and we may let $l\to\infty$ in the respective relations.
In Sec.\ \ref{SecIID}, the notion of 'asymptotically large' fractals
will be given a more precise meaning. 
Systems smaller than asymptotically large will be termed {\it finite} systems.
Asymptotically large systems are of interest in applications above all 
to astrophysical plasma-systems, where fractals may indeed be very large.

\subsection{The fractal scaling behaviour\label{SecIIB}}

\begin{figure}[h]
\resizebox{\hsize}{!}{\includegraphics{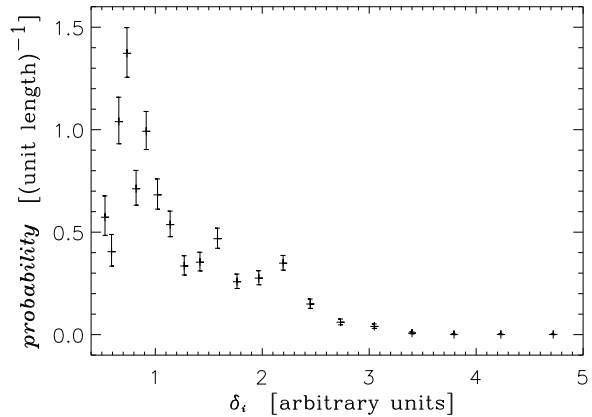}}
\caption{Histogram of the nearest neighbour distances $\delta_i$, for the set 
$F_3$ ($D_F=1.8$, $l=50$, $\delta=0.5$, see Sec.\ \ref{SecVIA1} and Table 
\ref{table1}).
\label{fig00} }
\end{figure}

Let us choose an arbitrary reference point $\vec x_i$ of the fractal,
somewhere in the interior (to neglect boundary effects).
Let $n_i(r)$ denote the number of points belonging to the fractal $F$
in the 3-dimensional sphere around $\vec x_i$ with radius $r$ ($r\leq l$,
with $l$ the radial size of the fractal). Since $F$ is a fractal with dimension
$D_F$, it is expected that 
\begin{equation}
n_i(r) = A_i\, r^{D_F^{(i)}} ,
\label{nra}
\end{equation}
with $A_i$ a constant (Eq.\ (\ref{nra}) is based on the {\it mass-scaling}
definition of fractal dimension, which is a common, practical definition 
of fractal dimension, see e.g.\ \cite{Mandelbrot1982}). Eq.\ (\ref{nra}) holds 
actually in the limit $r\to 0$, but in practice it is known
that the scaling behavior appears often already clearly at finite
$r$, and the limit $r\to 0$ is not feasible. It is also worthwhile
noting that Eq.\ (\ref{nra}) defines the local fractal dimension 
$D_F^{(i)}$, which may fluctuate with different reference
points $\vec x_i$, the more, the less numerous the points of the fractal are. 
The average of Eq.\ (\ref{nra}) over the whole fractal $F$ is 
yet well defined (else $F$ would in practice not be called a fractal).
In our applications, we 
are interested in statistical results, averaged over the entire fractal,
i.e.\ over all possible reference points $\vec x_i$, so that in the following 
we use a single scaling behaviour 
\begin{equation}
n(r) = A r^{D_F}
\label{nrA}
\end{equation}
everywhere, which corresponds to the average of the local $n_i(r)$ 
[Eq.\ (\ref{nra})] over $i$. 
The constant $A$ is determined as follows: With every point 
$\vec x_i$ of the fractal is associated a scale $\delta_i$ --- the distance
to the nearest neighbour ---, at which the local scaling behaviour 
$n_i(r)$ breaks down ($n_i(\delta_i)=1$, so that 
$n_i(r) = (r/\delta_i)^{D_F^{(i)}}$), which determines the constant
$A_i$ in Eq.\ (\ref{nra}), $A_i=(1/\delta_i)^{D_F^{(i)}}$. 
The scale $\delta$ introduced in Sec.\ \ref{SecIIA}, where the scaling breaks 
totally down, is understood as the minimum of all the $\delta_i$, 
$\delta:=\min_i[\delta_i]$. For the average
$n(r)$ of Eq.\ (\ref{nrA}), we have to use an average scale $\delta_\ast$ 
at which the scaling breaks down on the average (i.e.\ $n(\delta_\ast)=1$). 
In the examples of fractals we will introduce below, we find the distributions 
of the $\delta_i$ to be very asymmetric: they show a clear peak, but
exhibit a tail which extends to large 
$\delta_i$. Fig.\ \ref{fig00} shows a typical example of a histogram of the 
$\delta_i$ for the set $F_3$ which will be introduced below in 
Sec.\ \ref{SecVIA1}.
This particlular shape of the distribution of the $\delta_i$ has as a 
consequence that the arithmetic mean value of the $\delta_i$ is not 
representative of an average scale, it overestimates it. We therefore 
define $\delta_\ast$ as \textit{the most probable value} of the distribution 
of the $\delta_i$, determining it by a histogram of the $\delta_i$. The 
fractal scaling behaviour thus takes the form
\begin{equation}
n(r) = \left({r\over \delta_\ast}\right)^{D_F} ,
\label{nr}
\end{equation}
with $\delta \leq r\leq l$. 

Since the radial size of the the fractal is $l$, it follows that $n(l)$ is the 
total number of points $n_F$ of the fractal, or, with Eq.\ (\ref{nr}),
\begin{equation}
n_F = \left( l \over \delta_{\ast} \right)^{D_F} .
\label{nf}
\end{equation}
This relation is actually analogous to the case of non-fractal sets: 
If we sample for instance a 3--dimensional cube of side-length $l$ with
a resolution $\delta_\ast$, then we would obviously find 
$(l/\delta_\ast)^3$ points. Eq.\ (\ref{nf}) holds of course only for
points in the interior of the fractal, towards the edge it is biased by edge 
effects.

\subsection{Analytical derivation of the probability distribution $p_r$
of the random walk increments\label{SecIIC}}

From Eq.\ (\ref{nr}), it follows that the number of points $m(r)\Delta r$ 
of the fractal in a spherical shell around an interior point $\vec x_i$ with 
inner radius $r$ and radial thickness $\Delta r$ is 
$m(r)\Delta r = {d\over dr} n(r) \cdot \Delta r$, or
\begin{equation}
m(r)\Delta r = {D_F\over \delta_\ast}\, 
                           \left({r\over \delta_\ast}\right)^{D_F-1} 
                                      \,\Delta r .
\label{mr}
\end{equation}
With every point of the natural fractal is associated 
a cross-section $\rho^2\pi$, within which an approaching particle
gets into contact (interacts) with a point (elementary volume) of the 
fractal (see Sec.\ \ref{SecIIA}). The entire shell has thus a total 
cross-section
$s(r)\Delta r = m(r)\Delta r\cdot \rho^2\pi$, i.e.\ 
\begin{equation}
s(r)\Delta r = \rho^2\pi\, \frac{D_F}{\delta_\ast}\, 
                         \left(\frac{r}{\delta_\ast}\right)^{D_F-1}
                                 \, \Delta r .
\label{sr}
\end{equation}
In order this to hold, the different 
points of the fractal should not overlap or cover each other with their
cross sections. In the direction perpendicular to the radius $r$ this
is guaranteed by the fact that $\rho \leq \delta/2$, half the smallest 
separation distance of the points of the fractal, with
$\rho$ the cross-sectional radius. In the radial direction,
it is also guaranteed, as long as we let $\Delta r\leq \delta/2$,
so that also in radial direction we can be sure that no point is hidden
by the cross-sectional surface of another point in front of it. 

Assume now that a particle has started from $\vec x_i$ and has
traveled freely a distance $r$ into a random direction. The probability
$q_r \,\Delta r$ to hit 
the fractal in the spherical shell between $r$ and $r+\Delta r$
is the ratio of the total cross-section of the shell (the occupied
area), divided by the area of the shell,  
$q_r \,\Delta r= s(r)\Delta r/4\pi r^2$, or, with some rearrangements,
\begin{equation}
q_r\,\Delta r = 
\frac{D_F\rho^2}{4 \delta_\ast^3}\, 
\left(\frac{r}{\delta_\ast}\right)^{D_F-3}\,\Delta r   .
\label{qr}
\end{equation}

Our scope is to derive the probability $p_r \Delta r$ for a particle
to travel freely a distance $r$ and then to hit the fractal
in the spherical shell between $r$ and $r+\Delta r$, starting 
from an arbitrary point of $F$. To derive
this probability, we divide the interval $[\delta, r]$, which the particle 
travels freely, into a large number of small intervals 
of size $\delta r$: $[r_1,r_2]$, $[r_2,r_3]$, ..., $[r_{n-1},r_n]$,
with $r_1=\delta$, $r_n = r$, and $r_{i+1} - r_i =\delta r$ for all $i$
(the interval $[0,\delta]$ is free of points of $F$,
and thus has not to be taken into account, since there are no points 
of the fractal closer than $\delta$).
The probability 
not to hit the fractal in the intervals $[r_i,r_{i+1}]$ is $1-q_{r_i}\delta r$,
so that the probability not to hit the fractal in all the small intervals
up to $r$, and to hit
it finally in the interval $[r,r+\Delta r]$ is
\bea
p_r \Delta r &=& (1-q_{r_1}\delta r) \cdot (1-q_{r_2}\delta r)\cdot\,...\,  
                                                  \nonumber  \\
 & & \times\,...\,\cdot (1-q_{r_{n-1}}\delta r) \cdot q_r \Delta r  ,
\label{hulu}
\eea
or 
\begin{equation}
p_r \Delta r = 
\begin{cases}
  \prod_{i=1}^{n-1}(1-q_{r_i}\delta r)  \cdot q_r \Delta r , 
                                                  & \ {\rm for}\  n\geq 2 , \\
   q_{r_1} \Delta r ,                               & \ {\rm for}\  n=1 .\\ 
\end{cases}
\label{prdra}
\end{equation}
By defining 
\begin{equation}
\pi_r:=\prod_{i=1}^{n-1}(1-q_{r_i}\delta r)  ,
\label{pir}
\end{equation}
$p_r$ can be rewritten as 
\begin{equation}
p_r \Delta r = 
\begin{cases}
   \pi_r  \cdot q_r \Delta r ,           & \ {\rm for}\  n\geq 2  ,\\
   q_{r_1} \Delta r        ,           & \ {\rm for}\  n=1 .   \\ 
\end{cases}
\label{prdr}
\end{equation}

We have to evaluate the product $\pi_r$
in the limit $n\to \infty$, where the small intervals get infinitesimal.
First, we note that
\bea
\ln (\pi_r) &=& 
\ln \left( \prod_{i=1}^{n-1}(1-q_{r_i}\delta r)  \right)  \nonumber \\
&=&  \sum_{i=1}^{n-1} \ln \left(1-q_{r_i}\delta r  \right) .
\label{lnpir}
\eea
$q_{r_i}$ is always positive and bounded by 
${D_F\rho^2 \over 4 \delta_\ast^3}\, 
                     \left({\delta\over \delta_\ast}\right)^{D_F-3}$,
since $\delta \leq r \leq l$, and since
the exponent is negative ($D_F-3 < 0$). The term $q_{r_i}\delta r$
gets thus arbitrarily small for $n\to\infty$, since this implies
that $\delta r \to 0$ ($\delta r$ is something like $(r-\delta)/n$,
or, independent of $r$, $(l-\delta)/n$),
and we may thus use the approximation $\ln(1+x)\approx x$, for
$x<<1$. Eq.\ (\ref{lnpir}) thus becomes
\begin{equation}
\ln (\pi_r) \approx
  \sum_{i=1}^{n-1} -q_{r_i}\delta r  ,
\end{equation}
which for $\delta r \to 0$ may be considered as a standard expression
for the Riemann integral of $-q_r$, with limits $\delta$ and $r$,
\begin{equation}
\ln (\pi_r) =
 \int\limits_{\delta}^{r} -q_{r\prime} \, dr\prime  ,
\label{lnpirint}
\end{equation}
where due to the limit the approximation has become exact.

\subsubsection{The case $D_F\ne 2$\label{SecIIC1}}

Inserting for $q_r$ from Eq.\ (\ref{qr}) into Eq.\ (\ref{lnpirint}), and
solving for $\pi_r$, one finds that in the case $D_F \ne 2$
\begin{equation}
\pi_r^{(D_F \ne 2)} =
\exp\left[
- {D_F\rho^2 \over 4 \delta_\ast^{D_F}}\, 
{ \left( r^{D_F-2} -  \delta^{D_F-2}\right) \over D_F - 2 }
\right] .
\label{pira}
\end{equation}
The probability $p_r \Delta r= \pi_r\,q_r\,\Delta r$ (Eq.\ \ref{prdr}) for 
a particle to
start from a point of the fractal, to travel freely a distance $r$, and then
to hit the fractal in a layer of depth $\Delta r$ is thus, by
inserting Eqs.\ (\ref{qr}) and (\ref{pira}),
and by rearranging,  
\bea
p_r^{(D_F \ne 2)} \Delta r &=& 
\exp\left[ D_F \,\rho^2   
 { \left( \left( \frac{r}{\delta_\ast}\right)^{D_F-2} - 
       \left(\frac{\delta}{\delta_\ast}\right)^{D_F-2}\right) 
                 \over 4(2-D_F)\delta_\ast^2  }\right] \nonumber \\
& & \times \, { D_F \,\rho^2\over 4\delta_\ast^3 }  
             \left( \frac{r}{\delta_\ast} \right)^{D_F-3}
\cdot \Delta r ,
\label{pra}
\eea
where $\delta\leq r \leq l$.

Notably, the radial size $l$ of the fractal does not appear in the relation
for $p_r$: $l$ determines only the upper cut-off of $p_r$, it does not 
influence its shape. The size $l$ is thus an exterior parameter of the 
problem we study and can take any value between $\delta$ and infinity, 
without leading to any contradiction: as shown in App.\ \ref{AppA}, 
the normalization of $p_r$ never exceeds 1, whatever the value of $l$ 
is.

\subsubsection{The case $D_F = 2$\label{SecIIC2}}

In the case $D_F = 2$ we find from Eq.\ (\ref{qr}) and Eq.\ (\ref{lnpirint})
\begin{equation}
\pi_r^{(D_F = 2)} =
\exp\left[
- {D_F\rho^2 \over 4 \delta_\ast^{D_F}}\, \ln \frac{r}{\delta} 
\right] .
\label{pirb}
\end{equation}
Eqs.\ (\ref{prdr}), (\ref{qr}), and (\ref{pirb}) yield
\bea
p_r^{(D_F = 2)} \Delta r &=& 
\exp\left[ - {D_F \,\rho^2\over 4\delta_\ast^{D_F}}  
 \, \ln \frac{r}{\delta}  \right] \nonumber \\
& & \times \, { D_F \,\rho^2\over 4\delta_\ast^3 } 
       \left( \frac{r}{\delta_\ast} \right)^{D_F-3}
\cdot \Delta r   ,
\eea
which can be further rearranged to become 
\begin{equation}
p_r^{(D_F=2)} \Delta r = 
{ D_F \,\rho^2\over 4\delta_\ast^3 } 
        \left( \frac{\delta_\ast}{\delta} \right)^{-\frac{D_F \,\rho^2}{4\delta_\ast^2 }}
        \left( \frac{r}{\delta_\ast} \right)^{D_F-3-\frac{ D_F \,\rho^2}{4\delta_\ast^2}}
\cdot \Delta r ,
\label{prb}
\end{equation}
and is thus a pure power-law. Again, as in the case $D_F \ne 2$ 
[Eq.\ (\ref{pra})], the radial size $l$ of the 
fractal appears just as an upper limit for the allowed values of $r$.

\subsection{The rate for unaffected escape\label{SecIID}}

$\pi_r$ as defined in Eq.\ (\ref{pir}) is the probability not to hit 
the fractal at all in $[\delta,r]$ [see the explanation before 
Eq.\ (\ref{hulu})], so that $\pi_r\vert_{r=l}$ is obviously
the probability $\nu_{esc}$ not to hit the fractal at all, but to 
move unaffected by the fractal to the edge of the system and to
finally escape. From Eq.\ (\ref{pira}) 
we find that for $D_F\ne 2$, after slightly rearranging,
\bea
&&\nu_{esc}(D_F \ne 2) =       \nonumber \\
 &&\exp  \left[ \frac{ D_F \rho^2   
 \left( \left(\frac{l}{\delta_\ast}\right)^{D_F-2} - 
 \left(\frac{\delta}{\delta_\ast}\right)^{D_F-2}\right) }
                     {4(2-D_F)\delta_\ast ^2} \right] ,
\label{nuesca}
\eea
and in the case $D_F = 2$, from Eq.\ (\ref{pirb}),
\begin{equation}
\nu_{esc}(D_F = 2) =
\exp\left[
 - {D_F\rho^2 \over 4 \delta_\ast^{D_F}}\, \ln \left(\frac{l}{\delta}\right) 
\right] .
\label{nuescb}
\end{equation} 

Eqs.\ (\ref{nuesca}) and (\ref{nuescb}) imply that, depending mainly on 
the values of $D_F$, $l$, $\delta$, $\delta_\ast$, and $\rho$, there possibly 
is a finite rate for unaffected
escape, i.e.\ a finite fraction of the particles does not 'see' the fractal
and moves through the system without collisions
until it finally leaves.
Actually, for {\it finite} systems ($l<\infty$), there is {\it in any case} 
a finite rate of unaffected escape, which is the larger, the smaller the 
system size ($l$), the cross-sectional radius $\rho$, and the fractal 
dimension $D_F$ are.
For very large systems though, $\nu_{esc}$ settles to an asymptotic 
value, which corresponds to the lowest possible
escape rate for given $\rho$ and $D_F$.
As explained in Sec.\ \ref{SecIIA},
we will in the following call systems {\it asymptotically large} (in contrast
to {\it finite} systems),
if they are so large that $\nu_{esc}$ has practically
settled to its asymptotic value, and we will determine
$\nu_{esc}$ in their case by letting $l \to \infty$.

For {\it asymptotically large} systems ($l\to \infty$), we have the following 
cases, depending on the value of $D_F$:
\begin{itemize}
\item 
In the case $D_F>2$, we find from Eq.\ (\ref{nuesca})
\begin{equation}
\nu_{esc}(D_F>2, l\to \infty) = 0 ,
\end{equation}
so that all particles will collide with the asymptotically large fractal. 

\item 
In the case $D_F=2$, Eq.\ (\ref{nuescb}) yields
\begin{equation}
\nu_{esc}(D_F=2, l\to \infty) = 0 ,
\end{equation}
and again all particles collide with the asymptotically large fractal. 

\item For $D_F<2$, we have from Eq.\ (\ref{nuesca})
\begin{eqnarray}
&&\nu_{esc}(D_F<2,l\to\infty) \nonumber \\
&&= \exp \left[- \frac{D_F \rho^2}{4(2-D_F)\delta_\ast ^2} 
                \left(\frac{\delta}{\delta_\ast}\right)^{D_F-2} \right] ,
\end{eqnarray}
which is strictly smaller than $1$ (note that the argument of the 
exponential function is in any case negative and finite), so that there is a 
finite fraction $\nu_{esc}$ of particles which move through the system
without having any encounter along their path with the asymptotically 
large fractal, until they escape. 

\end{itemize}

\begin{figure}[h]
\resizebox{\hsize}{!}{\includegraphics{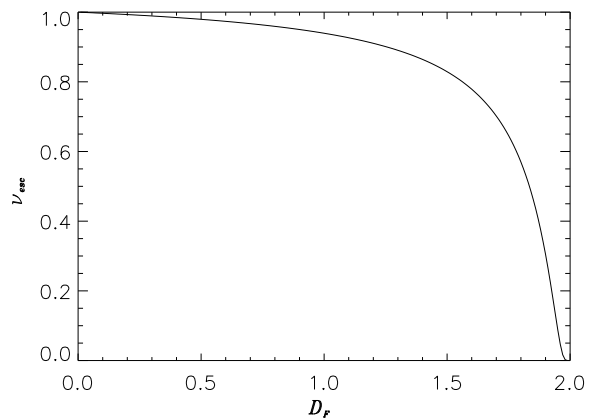}}
\caption{The escape rate $\nu_{esc}$ vs.\ the fractal dimension $D_F$, 
for $0<D_F<2$ and assuming asymptotically large systems ($l \to \infty$;
see Sec.\ \ref{SecIID}).
\label{fignu} }
\end{figure}

In Fig.\ \ref{fignu}, the rate of unaffected escape $\nu_{esc}$ is plotted 
against $D_F$ for $D_F < 2$, assuming asymptotically large systems 
($l\to\infty$): 
the escape rate is high, and only when $D_F$ approaches quite close $2$,  
the escape rate drops to low values.

It is to note that the particles which escape unaffected 
do not leave the system instantaneously, they remain 
in the system and move on a straight line path with their 
individual finite velocity, without ever colliding again
with the fractal, until they reach the edge of the system and leave.
In other words, the paths the escaping particles follow never
and nowhere intersect the fractal.
In the case of asymptotically large fractals, the time elapsing 
until an escaping particle reaches the edge
of the system may of course be considerable and much larger 
than the time for which the particles are tracked.

In Appendix \ref{AppA}, Eqs.\ (\ref{nuesca}) and (\ref{nuescb}) will 
be derived in an alternative way, and it will be shown that the possibly
finite rate of unaffected escape is related to the fact that $p_r$ is not 
necessarily normalized to one, it actually holds that 
\beq
\nu_{esc} = 1- \mu ,
\label{numu}
\eeq
where 
\begin{equation}
\mu := \int\limits_{\delta}^l p_r \, dr
\label{mu}
\end{equation}
is the normalization of $p_r$. The cases of finite escape 
rate (for asymptotically large ($l\to\infty$) as well for finite system size) 
correspond thus to the 
cases where $\mu < 1$, i.e.\ to the cases where $p_r$ is {\it defective}.

\subsection{Approximate forms of $p_r$\label{SecIIE}}

To determine possible approximate or asymptotic forms of $p_r$, we consider
the logarithmic derivative of $p_r$ [Eq.\ (\ref{pra})] for 
$D_F \ne 2$, 
 \begin{equation}
{ d\ln p_r^{(D_F \ne 2)}\over d\ln r }
=
-{D_F\,\rho^2 \over 4\delta_\ast^2} \left({r\over \delta_\ast}\right)^{D_F-2}
+ \left(D_F - 3 \right) .
\label{balu}
\end{equation}
The term $D_F-3$ stems from the power-law factor, and the
correcting term from the exponential factor in Eq.\ (\ref{pra}).
For $D_F<2$, the logarithmic slope asymptotically reaches $D_F-3$
for large $r$, being slightly distorted for small $r$, at most 
by the amount $(D_F\,\rho^2 /4\delta_\ast^2)\cdot(\delta/\delta_\ast)^{D_F-2}$ 
(for the smallest $r$, i.e.\ $r=\delta$). 
Hence, for $D_F < 2$, $p_r$ can be considered as an approximate power-law with 
index $D_F-3$, whose exact form is found from 
Eq.\ (\ref{pra}) on replacing $r$ in the exponential by its maximum possible 
value $l$,
\bea
p_r^{(a;\ D_F<2)} \Delta r &=& 
\exp\left[  \frac{D_F \rho^2    
 \left( \left({l\over \delta_\ast}\right)^{D_F-2} - 
       \left({\delta\over \delta_\ast}\right)^{D_F-2}\right) } 
                   {4(2-D_F)\delta_\ast^2} \right] \nonumber \\
& & \times \, { D_F \,\rho^2\over 4\delta_\ast^3 }  
      \left( r\over \delta_\ast \right)^{D_F-3}
\cdot \Delta r  .
\label{pr}
\eea
It follows that for $D_F < 2$ the second moments 
($\int r^2 p_r \, dr$) are infinite (the moments are dominated 
by the asymptotic, large $r$ regime), so that the random walks 
in the cases $D_F < 2$ are approximate realizations of {\it Levy-flights}:
for large $r$, the $p_r^{(a;\ D_F<2)}$ are of the same form as the 
Levy distributions, namely power-laws with index between $-3$ and $-1$
(see e.g.\ \cite{Hughes1995}), and
it is actually the large $r$ regime which causes the second moments
to diverge and the random walk statistics not to obey the Central Limit
Theorem. A characteristic difference to the Levy distributions is though
that the distributions $p_r^{(a;\ D_F<2)}$ are in any case 
defective, associated with a finite escape rate (Sec.\ \ref{SecIID}).

For $D_F>2$, the logarithmic slope in Eq.\ (\ref{balu}) is dominated
by the first term on the r.h.s., which increases in magnitude with 
increasing $r$, so that 
$p_r$ is decaying exponentially for large $r$, which implies 
that the second moments are finite, and the corresponding random walks are 
governed by the Central Limit Theorem.

The case $D_F=2$ is a pure power-law without
any approximation (Eq.\ \ref{prb}), the second moment is obviously infinite, 
and the random walk is an approximate realization of a Levy-flight, 
as are the cases $D_F<2$, defective 
though in the case of finite systems (see Sec.\ \ref{SecIID}).

\section{Continuous time random walk, generalized to the case of defective 
distributions: theory \label{SecIII}}

In order to determine the diffusive behavior of particles analytically, we 
follow the formalism of Continuous Time Random Walk (CTRW; see 
\cite{Montroll1965}) in the 
version of the velocity model (see \cite{Zumofen1993,Drysdale1998}). 
In this approach, it is assumed 
that each spatial walk increment $\vec r$ is performed in finite time $\tau$, 
where $r( \equiv \vert \vec r \vert )$ and $\tau$ are related through the velocity 
$v$ of the 
walker, which we assume to be arbitrary and constant. 
If we would not take into account the time spent in the 
jumps, then the mean square displacement we calculate below would be infinite 
in the cases where $D_F<2$, since the second moments of $p_r$ are 
infinite (Sec.\ \ref{SecIIE}), so that actually only the formalism of CTRW 
makes sense.

The connection between travel time $\tau$ spent in a jump and spatial 
increment $\vec r$ is expressed by 
the joint probability density $\psi(\vec r,\tau)$ to perform an unhindered 
walk-increment $\vec r$ in time $\tau$, which, in its simplest form, is 
\begin{equation}
\psi(\vec r,\tau) = p(\vec r) \delta(\tau -\vert \vec r \vert / v ) ,
\label{psirt}
\end{equation}
where the $\delta$-function just expresses the fact that a walk increment 
$\vec r$ takes time $\tau=\vert\vec r\vert / v$ to be performed 
($\delta(\tau -\vert \vec r \vert / v )$ is actually the conditional 
probability for the time spent in jump to equal $\tau$, given that the
jump length is $\vert \vec r\vert$). 
The spatial part $p(\vec r)$ of Eq.\ (\ref{psirt}) is the probability to make 
a jump $\vec r$, and it is 
is given through $p_r$ [Eq.\ (\ref{pra}), (\ref{prb}) or (\ref{pr})] as  
\begin{equation}
p(\vec r) \equiv p(\vert\vec r\vert ) = {p_r \over 4\pi r^2} .
\label{pvecr}
\end{equation}
(Note that $p_r$ is the probability to jump a distance $r$ 
into any direction, it is thus the marginal probability distribution of 
$p(\vec r)$, integrated over all directions,
$p_r = \int p(\vec r) \,d\sigma$, with 
$d\sigma = r^2 \sin\theta \,dr\,d\theta\,d\phi$ the usual surface element 
in spherical coordinates, so that $p_r = 4\pi r^2 p(\vert \vec r \vert)$ 
in the case where $p(\vec r)$ is isotropic.)

The formalism to determine the diffusive behaviour in the frame 
of CTRW for given jump- and flight-time distributions is presented
e.g.\ in \cite{Zumofen1993,Drysdale1998}. 
We have though to generalize this formalism in order to make it possible
to treat the case of possibly defective jump distributions.

\subsection{The propagator\label{SecIIIA}}

The basic quantity to be derived in order to determine the diffusive behaviour
is the so-called propagator $P(\vec r,t)$, the probability density for a particle
to be at position $\vec r$ at time $t$.
Thereto, 
we first have to determine the probability distribution $Q(\vec r, t)$ 
of the turning points (the points where the random walker changes 
direction), for which holds
\begin{equation}
Q(\vec r,t) = \int\!\! d^3r\prime \!\! \int\limits_0^t \!\!d\tau\,
Q(\vec r - \vec r\prime,t-\tau) \, \psi(\vec r\prime,\tau) 
+ \delta(t)\delta(\vec r)  .
\label{Qrt}
\end{equation}
This equation states that 
the probability to be at a turning point $\vec r$ at time $t$ equals 
the probability to be at the turning point $\vec r -\vec r\prime$ at time $t-\tau$, 
and to jump $\vec r\prime$ during time $\tau$, namely onto 
the turning point $\vec r$ exactly at time $t$. 
The second term on the r.h.s.\ explicitly 
takes the initial condition into account, assuming that all the random 
walkers start at the point $\vec r = 0$ at time $t=0$. 
In between turning points, the random walker is moving with constant velocity
$v$ on a straight line segment.
The probability $P(\vec r,t)$ to be at $\vec r$ at time $t$ is determined as 
\begin{equation} 
P(\vec r,t) = \int\!\! d^3r\prime \!\! \int\limits_0^t \!\! d\tau\,
Q(\vec r - \vec r\prime,t-\tau) \, \Phi(\vec r\prime,\tau)  ,
\label{Prt}
\end{equation}
where $\Phi(\vec r,\tau)$ is the probability to travel a distance  
$\vec r$ in time $\tau$, while making a jump of any length between 
$r \equiv \vert \vec r \vert$ and $\infty$, i.e.\ while either being on the 
way to the next turning point, or while moving unaffected on a path
leading to escape, 
\begin{eqnarray}
&&\Phi(\vec r,\tau) =: \Phi^{(c)}(\vec r,\tau) + \Phi^{(e)}(\vec r,\tau) 
                                \nonumber \\
&=&  \delta(\tau - \vert \vec r \vert / v ) 
  \left[ \frac{1}{4\pi r^2}
  \int\limits_{\vert \vec r\prime\vert \geq \vert \vec r\vert}
    \!\!              dr\prime \, p_{r\prime} 
              + \frac{\nu_{esc}}{4\pi r^2}        \right] 
\label{Phirt} 
\end{eqnarray} 
with $p_r$ from Eq.\ (\ref{pra}), (\ref{prb}) or (\ref{pr}), and where on 
the r.h.s.\ we identify the first term 
as the 'collisional' term $\Phi^{(c)}(\vec r,\tau)$ and the second term 
as the 'escape' term $\Phi^{(e)}(\vec r,\tau)$. 
The appearance of the 
'escape' term is a consequence of the possible defectiveness of $p_r$,
if $p_r$ is normalized to one ($\mu=1$) then this term disappears 
($\nu_{esc}=1-\mu=0$, see Sec.\ \ref{SecIID}). It takes into account the
particles which have started from a turning point and are moving 
unaffected until they escape, not colliding anymore with the fractal 
on their path.
Eq.\ (\ref{Phirt}) holds in the range $\delta \leq r \leq \infty$. In the 
range $0\leq r \leq \delta$, all the particles move unhindered, either they 
are on a unaffected escape path or they are on the way to their next turning 
point, since there are no points of the fractal closer than $\delta$ 
(see Sec.\ \ref{SecIIA}), so that for $r\leq \delta$
\beq
\Phi(\vec r,\tau) =: \Phi^{(0)}(\vec r,\tau) 
=  \delta(\tau - \vert \vec r \vert / v) 
              \cdot \frac{1}{4\pi r^2}    .
\label{Phi00}
\eeq

With the description of $\Phi(\vec r,\tau)$, it is now clear that  
Eq.\ (\ref{Prt}) expresses the fact that a particle is (i) either at a turning 
point ($\vec r\prime =0$, $\tau = 0$), or (ii) has started from a turning 
point (at $\vec r - \vec r\prime$, $t-\tau$) and is now traveling  
towards 
its next turning point, not yet having reached it, though, 
($\tau > 0$, $\vec r\prime \ne 0$)
and passes by the the point $\vec r$ at time $t$, 
or (iii) the particle has started from a turning point 
(at $\vec r - \vec r\prime$, $t-\tau$) and moves
unaffected on an escape path ($\tau > 0$, $\vec r\prime \ne 0$), 
passing by the the point $\vec r$ at time $t$.

Eq.\ (\ref{Prt}) is an integral equation for $P(\vec r,t)$, with 
$\psi(r,\tau)$ given, together with the auxiliary integral equation
for $Q(\vec r,t)$ (Eq.\ \ref{Qrt}).
As pointed out in Sec.\ \ref{SecIIA},
in every reasonable application
the system is finite, i.e.\ the fractal is of finite size 
($l<\infty$), and the particles definitely leave the 
region occupied by the fractal when they have reached a distance 
from the origin equal to the radial size of the fractal. This
implies that the spatial integrals in Eqs.\ (\ref{Qrt}) and (\ref{Prt}) 
are actually over a finite
range (equal to the linear size of the fractal), and somewhat involved methods
have to be used to solve the integral equations 
(see e.g.\ \cite{Drysdale1998} for a 
study of these combined integral equations for a finite system in 
1-dimensional space and in the non-defective case). 
Here, we simplify the problem 
by assuming that the fractal is very large, so that assuming an infinite 
system size $l$ should give a good impression of the diffusive behavior. The 
finite system size acts merely as an upper cut-off for the possible
range of values of the distances from the origin that particles travel. 
Since we again let $l\to \infty$, as in Sec.\ \ref{SecII},
we can formally identify the very large systems we have in mind
here with the asymptotically large systems introduced in Sec.\ \ref{SecII}.
For asymptotically large systems now ($l\to \infty$), the combined integral 
equations 
(\ref{Prt}) and (\ref{Qrt}) are most easily 
solved by Fourier-transforming in space ($\vec r \to \vec k$) and 
Laplace-transforming in time ($t \to s$), applying the respective
Laplace and Fourier convolution-theorems (see e.g.\ \cite{Morse}), 
which yields
\begin{equation}
P(\vec k,s) = {\Phi(\vec k, s) \over 1-\psi(\vec k,s)}
\label{Pks}
\end{equation}
Eq.\ (\ref{Pks}) is formally identical to the non-defective case
(see \cite{Zumofen1993}), it is to note though that $\Phi$ is defined 
in a different, generalized way.

\subsection{The diffusive behaviour\label{SecIIIB}}

The mean square displacement 
\begin{equation}
\langle \vec r^2(t) \rangle := \int\limits \vec r^2 P(\vec r,t) \, d^3r
\label{rsqtdef}
\end{equation}
can straight forwardly be shown to equal to
\begin{equation}
\langle \vec r^2(t) \rangle 
= - {d^2\over d\vec k^2} P(\vec k,t) \big\vert _{\vec k = 0} 
\label{rsqt}
\end{equation}
(by inserting the definition of Fourier transform). 
To calculate $\langle \vec r^2(t) \rangle$ through Eqs.\ (\ref{Pks}) and (\ref{rsqt})
analytically in the Secs.\ \ref{SecIV} and \ref{SecV},
we will make the following assumptions: (i) $s<<1$ (since we are 
interested in the case of $t \to \infty$), 
(ii) $\vert\vec k\vert << 1$ (corresponding to asymptotically large systems, 
$l\to \infty$),
and (iii) $\vert\vec k\vert << s$ (since, according to 
Eq.\ (\ref{rsqt}), we will at the end set $\vec k = 0$).

\subsection{The expected number of jumps in a given time-interval 
              \label{SecIIIC}}

Since the escape rate can be finite, it will be interesting to know
how many times a particle collides on the average with the fractal before 
it escapes.
We determine thus in this section a relation for 
the expected number of jumps $\langle N(t)\rangle$ in a given time 
interval $[0,t]$. This relation is in principle given e.g.\ in \cite{Hughes1995},
we have though to clarify whether the relation in \cite{Hughes1995}
is applicable to the cases of defective jump distributions.

We determine first 
the distribution of travel times $\varphi(t)$, i.e.\ the distribution of the 
times spent in a single jump, as the marginal distribution of $\psi(\vec r,t)$
(Eq.\ \ref{psirt})
\begin{equation}
\varphi(\tau) := \int \psi(\vec r,\tau) \,d^3r .
\label{phitau}
\end{equation}
Concerning the normalization of $\varphi(\tau)$, we note that 
\beq
\int_{\delta/v}^\infty \varphi(\tau) \,d\tau 
           = \int_\delta^\infty p_r \,dr 
           = \mu  
\label{phitaunorm}
\eeq 
(see App.\ \ref{AppB3}),
the normalization of $\varphi(\tau)$ is thus identical to the
normalization of $p_r$, which we defined to be $\mu$ in Eq.\ (\ref{mu}). 
The distribution of travel times $\varphi(\tau)$ is thus defective ($\mu < 1$)
in the cases where the distribution of jump increments $p_r$ is defective.

The probability $\varphi_n(t)$ for the $n$th jump 
to take place at time $t$ is recursively determined by 
\begin{equation}
\varphi_n(t) = \int\limits_0^t \varphi(\tau) \varphi_{n-1}(t-\tau) \, d\tau ,
\label{laco1}
\end{equation}
i.e.\ if the $(n-1)$th jump took place at time $t-\tau$ and
was followed by a jump of duration $\tau$, then the $n$th jump
takes place at time $t$.
Laplace-transforming yields $\varphi_n(s) = \varphi(s) \varphi_{n-1}(s)$ (through
the Laplace convolution theorem),
and if we iterate, we are led to
\begin{equation}
\varphi_n(s) = \varphi(s)^n .
\label{b2}
\end{equation}

The probability ${\rm prob}[N(t) = n]$ that the number of jumps $N(t)$ made in the 
time interval $[0,t]$ equals a given number $n$ is given as
\begin{equation} 
{\rm prob}[N(t) = n] =  \int\limits_0^t \varphi_n(t\prime) \Xi(t-t\prime) \, dt\prime ,
\label{b3}
\end{equation}
with $\Xi(t-t\prime)$ the probability to make a jump of duration at least 
$t-t\prime$. Eq.\ (\ref{b3}) states that the $n$th jump took place at time 
$t\prime$, and the subsequent jump took longer than $t-t\prime$, so that 
there was no subsequent jump completed in $[t\prime ,t]$. 
$\Xi(t)$ is determined as 
\begin{eqnarray}
\Xi(t) =
  \int\limits_{t}^\infty \varphi(\bar{t}) \,d\bar{t} + \nu_{esc}  ,
\label{b4}
\end{eqnarray}
where the first term on the r.h.s.\ is the probability that a particle
makes a jump of duration $t$ or longer, and the second term is the
probability that a particle moves unaffected on a path leading to escape, 
having thus an infinite travel time. Using $\mu=\int_o^\infty \varphi(t)\,dt$ 
[see Eq.\ (\ref{phitaunorm})], we can write
Eq.\ (\ref{b4}) as
\begin{eqnarray}
\Xi(t)  
&=& \mu - \int\limits_{0}^{t} \varphi(\bar{t}) \,d\bar{t} + \nu_{esc} 
                                                       \nonumber \\
&=& 1 - \int\limits_{0}^{t} \varphi(\bar{t}) \,d\bar{t} ,
\label{b5}
\end{eqnarray}
where we have used the fact that $\mu + \nu_{esc} = 1$ [Eq.\ (\ref{numu})]. 
The Laplace transform of Eq.\ (\ref{b5}) is  
\begin{equation}
\Xi(s) = {1\over s}(1 - \varphi(s)) ,
\label{b66}
\end{equation}
and Laplace-transforming Eq.\ (\ref{b3}) yields
\begin{eqnarray}
{\rm prob}[N(s) = n] &=& \varphi_n(s) \Xi(s) \nonumber \\
&=&  \varphi(s)^n \Xi(s) ,
\label{b6}
\end{eqnarray}
where we have inserted also Eq.\ (\ref{b2}), and on replacing $\Xi(s)$
by Eq.\ (\ref{b66}), we find
\begin{equation}
{\rm prob}[N(s) = n] = \varphi(s)^n {1\over s}(1 - \varphi(s)) .
\label{b7}
\end{equation}

The expected number of jumps $\langle N(t) \rangle$ in the time-interval 
$[0,t]$ follows from the definition of expectation value,
\begin{equation}
\langle N(t) \rangle = \sum\limits_{n=0}^\infty n\cdot {\rm prob}[N(t) = n] ,
\end{equation}
which in Laplace space becomes, when also inserting Eq.\ (\ref{b7}),
\begin{eqnarray}
\langle N(s) \rangle &=& \sum\limits_{n=0}^\infty n\cdot {\rm prob}[N(s) = n] 
                                                     \nonumber  \\
&=& {1\over s}(1 - \varphi(s)) \sum\limits_{n=0}^\infty n\cdot \varphi(s)^n  .
\end{eqnarray}
The sum can be evaluated by using the relations 
$\sum_{n=0}^\infty nx^n = x (d/dx) \sum_{n=0}^\infty x^n$ and 
$\sum_{n=0}^\infty x^n = 1/(1-x)$, 
which finally yields 
\begin{equation}
\langle N(s) \rangle = \frac{\varphi(s)}{s(1-\varphi(s))} .
\label{Ns}
\end{equation}

It thus turned out that the expression for $\langle N(s) \rangle$ 
in the defective case is identical to 
the relation for the case where $\varphi(t)$ is normalized to one
(see e.g.\ \cite{Hughes1995}).
The essential modification in the derivation for the defective case was 
the addition of the term $\nu_{esc}$ in Eq.\ (\ref{b4}).

Contrary to the relations which determine $\langle \vec r^2(t) \rangle$
[mainly Eq.\ (\ref{Pks})], the formula for $\langle N(s) \rangle$
[Eq.\ (\ref{Ns})] is valid also in the case of finite systems ($l < \infty$):
The Laplace convolution-theorem we used to solve the integral equations 
(\ref{laco1}) and (\ref{b3}) is applicable to convolutions over finite 
intervals, 
contrary to the Fourier convolution-theorem used in Sec.\ \ref{SecIIIA}, 
which demands infinite integration intervals in order to be applicable; 
see e.g.\ \cite{Morse}.

\section{Application of the CTRW formalism to the case $D_F <2$
\label{SecIV}}

\subsection{Diffusion for $D_F <2$ \label{SecIVA}}

We analyze the diffusive behavior for
the case $D_F<2$, where it had been shown in Sec.\ \ref{SecIIE} that the random
walk is of the type of defective Levy-flights. 
We will use the asymptotic power-law form $p_r^{(a;\ D_F<2)}$ for $p_r$ 
(Eq.\ \ref{pr}), writing for conciseness
\begin{equation}
p_r = C \, r^{D_F-3} ,
\label{prC}
\end{equation}
where $C$ summarizes the constant pre-factors in Eq.\ (\ref{pr}).
Eq.\ (\ref{prC}) implies through Eqs.\  (\ref{psirt}) and (\ref{pvecr})
for the joint probability distribution of jump increments and travel times
\begin{equation}
\psi(\vec r,\tau) = \frac{C}{4\pi}\, r^{D_F-5} 
                       \delta(\tau - \vert \vec r \vert / v ) .
\label{psirtC}
\end{equation}

By assuming that the system is asymptotically large ($l\to \infty$), 
so that formalism developed in Secs.\ \ref{SecIIIA} and \ref{SecIIIB} 
can be applied, the diffusive behaviour is determined through 
Eqs.\ (\ref{Pks}) and (\ref{rsqt}).
We need thus the Fourier- Laplace-transforms of $\psi(\vec r,t)$ 
[Eq.\ (\ref{psirtC})] and $\Phi(\vec r,t)$ [Eqs.\ (\ref{Phirt}) and 
(\ref{Phi00})]. 
The way we calculate the Fourier- and Laplace-transforms, also in the 
subsequent sections, with the conditions $s<<1$ and $k<<s$ 
(see Sec.\ \ref{SecIIIB}) is described in App.\ \ref{AppB}:

The Fourier-Laplace-transform of $\psi(\vec r,t)$ [Eq.\ (\ref{psirtC})]
for $D_F>1$ is
\bea
\psi(\vec k,s)^{(D_F>1)} &\approx&
\mu  
- Cv^{D_F-2}\Gamma(D_F-1) \cdot s^{2-D_F} \nonumber \\
&&- {1\over 6} k^2 C v^{D_F} \Gamma(D_F) \cdot s^{-D_F}  ,
\label{psiksa}
\eea 
and for $D_F<1$ it is
\bea
\psi(\vec k,s)^{(D_F<1)} &\approx&
\mu - \langle T \rangle \cdot s          \nonumber \\
&&- {1\over 6} k^2 C v^{D_F} \Gamma(D_F) \cdot s^{-D_F}  ,
\label{psiksb}
\eea 
with $\Gamma(.)$ Euler's Gamma function, 
$\mu$ the normalization of $p_r$ [Eq.\ \ref{mu}],
and $\langle T \rangle$ the expectation 
value of the time spent in a single jump, defined in Eq.\ (\ref{Tmean}) below.

$\Phi(\vec r, \tau)$ [Eqs.\ (\ref{Phirt}) and (\ref{Phi00})] consists of three
parts: $\Phi^{(c)}(\vec r, \tau)$  is determined through 
Eqs. (\ref{prC}) and (\ref{Phirt}) as
\begin{equation}
\Phi^{(c)}(\vec r, \tau) = 
{C\over 4\pi (2-D_F)}\, r^{D_F-4} \, \delta(\tau - \vert\vec r\vert / v),
\label{ph1}
\end{equation}
whose Fourier-Laplace-transform for $D_F>1$ is (see App.\ \ref{AppB})
\bea
\Phi^{(c)}(\vec k, s)^{(D_F>1)} &\approx&
\frac{C v^{D_F-1} \Gamma(D_F-1)}{2-D_F}  \cdot s^{1-D_F} \nonumber \\
-&k^2& \frac{C v^{D_F+1} \Gamma(D_F+1)}{6(2-D_F)}   \cdot s^{-D_F-1}  ,
\label{Phiksa}
\eea
and for $D_F<1$ it becomes
\bea
\Phi^{(c)}(\vec k, s)^{(D_F<1)} &\approx&
\frac{C v^{D_F-1}}{(2-D_F)(1-D_F)}  \left(\delta\over v\right)^{D_F-1} 
                                                                 \nonumber \\
&&- \frac{C v^{D_F-1} \Gamma(D_F)}{2-D_F}   \cdot s^{1-D_F} \nonumber \\
-&k^2& \frac{C v^{D_F+1} \Gamma(D_F+1)}{6(2-D_F)}  \cdot s^{-D_F-1}  .
\label{Phiksb}
\eea

The Fourier-Laplace-transform of $\Phi^{(e)}(\vec r,\tau)$
[Eq.\ (\ref{Phirt})] is given as (see App.\ \ref{AppB})
\begin{equation}
\Phi^{(e)}(\vec k,s) = \nu_{esc} v \Gamma(1) \, s^{-1} 
     - k^2 \frac{\nu_{esc}v^3 \Gamma(3)}{6}  \, s^{-3} .
\label{Phikse}
\end{equation}

Last, the Fourier-Laplace transform of 
$\Phi^{(0)}(\vec r,\tau)$ [Eq.\ \ref{Phi00})] is (see App.\ \ref{AppB})
\beq
\Phi^{(0)}(\vec k,s) = a_1^{(0)} - a_2^{(0)} \cdot s
\label{Phiks0}
\eeq
with $a_1^{(0)}$, $a_2^{(0)}$ finite constants.

Inserting $\psi(\vec k,s)$ [Eqs.\ (\ref{psiksa}), (\ref{psiksb})] and 
$\Phi(\vec k,s)\equiv \Phi^{(c)}(\vec k,s) +\Phi^{(e)}(\vec k,s) 
                                                +\Phi^{(0)}(\vec k,s)$ 
[Eqs.\ (\ref{Phiksa}), (\ref{Phiksb}), (\ref{Phikse}) and (\ref{Phiks0})]
into Eq.\ (\ref{Pks}), 
differentiating $P(\vec k,s)$ according to Eq.\ (\ref{rsqt}),  
setting thereafter $\vec k$ zero, we find, neglecting the constants,
keeping only the leading terms in $s$ for $s\to 0$, and noting that 
$\mu \ne 1$, 
\begin{equation}
<\vec r^2(s)> \sim {1\over s^3} , \ \ \ \ {\rm for}\ 0 < D_F < 2  .
\label{r2s}
\end{equation}
The $D_F$-dependence (through $\psi(\vec k,s)$ and 
$\Phi(\vec k,s)$) has disappeared in the 
limit $s\to 0$, the behaviour is actually dominated by the 
$D_F$-independent escape term $\Phi^{(e)}(\vec k,s)$.

Since Eq.\ (\ref{r2s}) holds only for $s\to 0$, the direct 
Laplace back-transformation is not defined, and
we have to use the Tauberian theorems (see e.g.\ \cite{Feller1971}), 
which yield for $t$ large 
\begin{equation}
<\vec r^2(t)>  \sim t^2,    \ \ \ \ {\rm for}\ 0 < D_F < 2  .
\label{r2t}
\end{equation}
The diffusion is thus in any case anomalous, namely enhanced,
of the super-diffusive, ballistic type.

\subsection{$D_F<2$: the expected number of collisions \label{SecIVB}}

Since the escape rate $\nu_{esc}$ is in any case finite for $D_F<2$ 
(Sec.\ \ref{SecIID}), it is of interest 
to know how many times a particle collides on the average with the fractal 
before it escapes. Thereto,
we determine the expected number of jumps $\langle N(t)\rangle$ performed 
by a particle in the time interval $[0,t]$. According to Sec.\ \ref{SecIIIC}, 
we first have to determine $\varphi(\tau)$, which through Eqs.\ (\ref{phitau})
and Eq.\ (\ref{psirtC}) we find to be
\begin{equation}
\varphi(\tau) = C v^{D_F-2} \tau^{D_F-3}
\label{phitau1}
\end{equation} 
for $\tau \geq \delta /v$ (the minimum jump length is $\delta$, 
see Sec.\ \ref{SecIIA}).
 
For $D_F<1$, the Laplace transform of $\varphi(t)$ is (see App.\ \ref{AppB5})
\begin{equation}
\varphi(s) \approx \mu - \langle T \rangle \, s ,
\end{equation}
with $\mu$ the normalization of $\varphi(\tau)$ (see Eq.\ \ref{phitaunorm}),
and where $\langle T \rangle$ is the expectation value of the the time spent 
in a jump, 
\begin{equation}
\langle T \rangle = \int_{\delta/v}^\infty \tau \, \varphi(\tau)\, d\tau .
\label{Tmean}
\end{equation}
For $D_F>1$, the Laplace transform of $\varphi(t)$ becomes
\begin{equation}
\varphi(s) \approx \mu - C v^{D_F-2}\, \Gamma(D_F-1)\, s^{2-D_F} ,
\end{equation} 
the second term diverges for $s \to 0$, implying that the expected 
flight time $\langle T \rangle$ is infinite (see App.\ \ref{AppB}).

\begin{figure}[h]
\resizebox{\hsize}{!}{\includegraphics{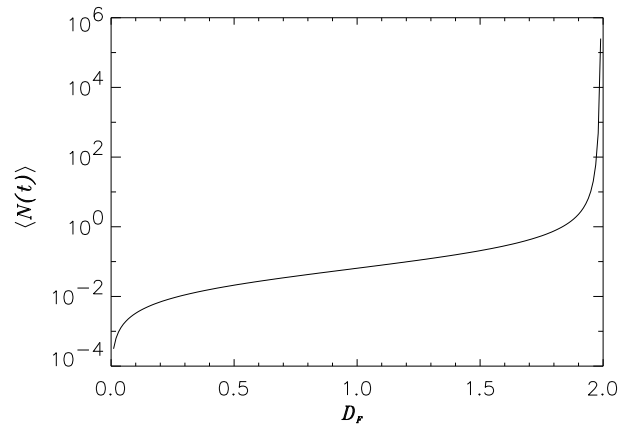}}
\caption{The expected number of collisions $\langle N(t) \rangle$ vs.\ 
fractal dimension $D_F$, for $0<D_F<2$, assuming
asymptotically large systems ($l \to \infty$) and large times (see Sec.\ 
\ref{SecIVB}).
\label{figND} }
\end{figure}

Inserting into Eq.\ (\ref{Ns}), we find, when keeping only the leading terms
in $s$ and noting that $\mu \ne 1$,
\begin{equation}
\langle N(s)\rangle \sim 
{\mu \over 1- \mu}{1\over s} ,  \ \ \ \ \ \ \  {\rm for}\ 0 < D_F < 2,
\label{NsNs}
\end{equation}
--- as in the case of $\langle \vec r^2(s)\rangle$ [Eq.\ (\ref{r2s})], 
$\langle N(s)\rangle$ is independent
of $D_F$, due to the fact that the normalization $\mu < 1$, i.e.\ the  
finite rate of unaffected escape $\nu_{esc}$ [$=1-\mu$, see Eq.\ (\ref{numu})] 
dominates the behaviour.
From Eq.\ (\ref{NsNs}), the Tauberian theorems yield for the back-transform
\begin{equation}
\langle N(t)\rangle \sim 
{\mu \over 1- \mu} \delta(t)  ,  \ \ \ \ \ \ \  {\rm for}\ 0 < D_F < 2 .
\end{equation}
The expected number of jumps is therewith constant,
it does not increase with time anymore for large enough times.
In Fig.\ \ref{figND}, we show $\langle N(t)\rangle$
as a function of the dimension $D_F$ ($D_F<2$) for asymptotically large systems
($l\to\infty$) and large times such that $\langle N(t)\rangle$ has settled
to its expected value.
Obviously, particles do very inefficiently interact with fractals of 
dimension below 2, they 
almost do not see the fractals, and only if the dimension approaches 
quite close the value 2, collisions with the fractal become numerous and
important. For finite systems, the escape rate $\nu_{esc}$
is still larger (see Sec.\ \ref{SecIID}), and collisions with the fractal get 
even more rare.

\section{Application of the CTRW formalism to the case $D_F>2$ \label{SecV}}

\subsection{Diffusion for $D_F >2$ \label{SecVA}}

For $D_F>2$, $p_r$ cannot be approximated by a power-law 
(see Sec.\ \ref{SecIIE}), we have to 
keep the full form of $p_r$ in Eq.\ (\ref{pra}), and
the random walk is governed by the Central Limit Theorem,
since all the moments of $p_r$ are finite. 
We thus expect diffusion to be normal, a theoretical
expectation we have to confirm in the following. 

We assume the system to be asymptotically large ($l\to\infty$), so that 
we can apply the formalism of Secs.\ \ref{SecIIIA} and \ref{SecIIIB},
and moreover it follows that
$\nu_{esc}=0$ and $\mu=1$ (see Sec.\ \ref{SecIID}).
In order to determine $\langle\vec r^2(t)\rangle$ through Eq.\ (\ref{rsqt}),
we have first to determine the joint distribution for walk increments and 
flight times $\psi(\vec r,t)$ [Eq.\ (\ref{psirt})], and the distribution 
$\Phi(\vec r,t)$ to make a jump of at least length $r$ [Eqs.\ (\ref{Phirt})
and (\ref{Phi00})].
For convenience, we write $p_r$ [Eq.\ (\ref{pra})] in the form 
\begin{equation}
p_r = C \exp\left[-\beta r^{D_F-2}\right] r^{D_F-3} ,
\label{prr}
\end{equation}
where all the constants in Eq.\ (\ref{pra}) are incorporated in the
constants $C$ and $\beta$ in an obvious manner. Eqs.\ 
(\ref{prr}), (\ref{psirt}), and (\ref{pvecr})
imply that 
\begin{equation}
\psi(\vec r,t) = \frac{C}{4\pi} \exp\left[-\beta r^{D_F-2}\right] 
                                          r^{D_F-5} \delta(t-r/v) .
\label{psiDgt2}
\end{equation}
The Fourier-Laplace transform of $\psi(\vec r,t)$ is found to be
(see App.\ \ref{AppB})
\begin{equation}
\psi(\vec k,s) \approx 
 \mu - \langle T\rangle \cdot s  - 
       \frac{1}{6} k^2 v^2 \left( \langle T^2\rangle - \langle T^3\rangle \cdot s \right) ,
\end{equation}
where $\mu$ is the normalization of $p_r$ and, since we assume the system to 
be asymptotically large ($l\to\infty$), we have $\mu=1$ (Sec.\ \ref{SecIID}).
The $\langle T^n \rangle < \infty$ are constants, whose exact values are not 
relevant for our purposes, here (they are actually the moments of the
distribution $\varphi(\tau)$ which is introduced below in Sec.\ \ref{SecVB},
see App.\ \ref{AppB}). 

The collisional part $\Phi^{(c)}(\vec r,t)$ of $\Phi(\vec r,t)$ is given 
through Eqs.\ (\ref{Phirt}) and (\ref{prr}),
\begin{eqnarray}
\Phi^{(c)}(\vec r,t)    
&=& 
\frac{C}{4\pi\beta(D-2)}  \exp\left[-\beta r^{D_F-2}\right] r^{-2}  
                                                           \nonumber \\
&& \qquad\qquad\qquad\qquad \times \delta(t-r/v)   .
\label{PhicDgt2}
\end{eqnarray}
Fourier-Laplace-transforming $\Phi^{(c)}(\vec k,s)$ yields 
(see App.\ \ref{AppB})
\begin{equation}
\Phi^{(c)}(\vec k,s) \approx b_1 - s\cdot b_2 - k^2 (b_3 - s\cdot b_4)  ,
\end{equation}
with the $b_i<\infty$ constants.

The 'escape' term $\Phi^{(e)}(\vec r,t)$ of $\Phi(\vec r,t)$ 
[Eq.\ (\ref{Phirt})] 
is zero for asymptotically large systems (since $\nu_{esc}=0$).

$\Phi^{(0)}(\vec r,t)$ [Eq.\ (\ref{Phi00})] is independent of $D_F$,
so that its Laplace-Fourier transform is given by Eq.\ (\ref{Phiks0}).

Having determined $\psi(\vec k,s)$ and $\Phi(\vec k,s) = 
\Phi^{(c)}(\vec k,s) + \Phi^{(0)}(\vec k,s)$,
we can turn to the determination of 
$\langle \vec r^2(s)\rangle$ through Eq.\ (\ref{rsqt}).
For the asymptotically large systems ($l \to \infty$), which we consider here, 
it holds $\mu=1$, so that the leading term in Eq.\ (\ref{rsqt}) 
for $s\to 0$ is  
\begin{equation}
\langle \vec r^2(s)\rangle \sim \frac{1}{s^2}  ,
\end{equation}
and the Tauberian theorems yield the back-transform
\begin{equation}
\langle \vec r^2(t)\rangle \sim t
\label{r2tb}
\end{equation}
for large times, i.e.\  diffusion is normal, as it is expected
from the Central Limit Theorem.

\subsection{$D_F>2$: the expected number of collisions \label{SecVB}}

According to Eq.\ (\ref{phitau}) and Eq.\ (\ref{prr}), the distribution 
$\varphi(\tau)$ of times spent in a jump is
\begin{equation}
\varphi(\tau) = C v^{D_F-2} \exp\left[-\beta v^{D_F-2}\tau^{D_F-2}\right] 
                                              \tau^{D_F-3} ,
\label{phitauDgt2}
\end{equation}
and for its Laplace-transform we find (see App.\ \ref{AppB5})
\begin{equation}
\varphi(s) \approx \mu - s \, \langle T \rangle ,
\end{equation}
where $\mu$ is the normalization of $\varphi(\tau)$ 
[see Eq.\ (\ref{phitaunorm})],  
and $\langle T \rangle$ the expected time spent in a single jump
[defined as in Eq.\ (\ref{Tmean})]. 

To determine $\langle N(s) \rangle$ through Eq.\ (\ref{Ns}), we discern 
between asymptotically large and finite systems:
For asymptotically large systems ($l \to \infty$), we have $\nu_{esc} = 0$ 
and $\mu=1$ (see Sec.\ \ref{SecIID}), 
and, keeping only the leading terms for $s\to 0$, 
Eq.\ (\ref{Ns}) yields  
\begin{equation}
\langle N(s) \rangle \sim  \frac{1}{\langle \tau \rangle} \frac{1}{s^2}  ,
\end{equation}
so that by the Tauberian theorems the back-transform is
\begin{equation}
\langle N(t) \rangle \sim \frac{t}{\langle \tau \rangle}
\end{equation}
--- for large times, the number of jumps is just the time
divided by the expected time a walker spends in a single jump.
This is a consequence of the Central Limit Theorem.

In the case of finite systems, the escape
rate is finite, $\nu_{esc} > 0$, so that $\mu \ne 1$ (Sec.\ \ref{SecIID}), 
and the leading term for $s\to 0$ in Eq.\ (\ref{Ns}) is 
\begin{equation}
\langle N(s) \rangle \sim \frac{\mu}{1-\mu} \frac{1}{s} .
\end{equation}
By using the Tauberian theorems, we find
\begin{equation}
\langle N(t) \rangle \sim \frac{\mu}{1-\mu} \delta(t) ,
\label{Ntd}
\end{equation}
the expected number of collisions is constant for large times, 
there is a finite, $\mu$-dependent, average number of collisions, after 
which a particle 
does not interact with the fractal anymore and moves unaffected
until it escapes.

\begin{figure}[h]
\resizebox{\hsize}{!}{\includegraphics{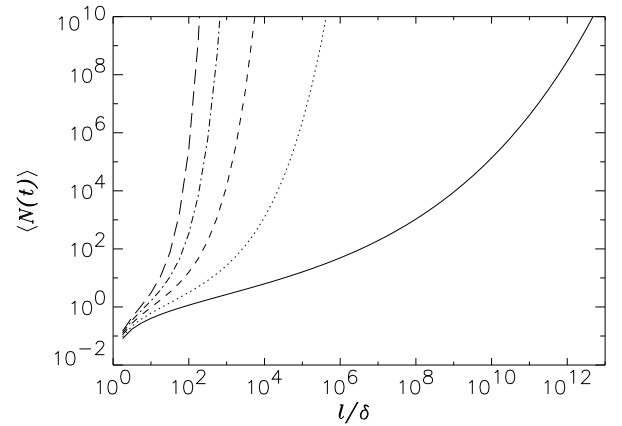}}
\caption{The expected number of collisions $\langle N(t) \rangle$ 
[Eq.\ (\ref{Ntd})] vs.\ 
the scaling range $l/\delta$ of fractals for large times, and 
for the cases $D_F=2.1$ (solid), $D_F=2.3$ (dotted), $D_F=2.5$ (short dash),
$D_F=2.7$ (dash-dot), and $D_F=2.9$ (long dash); see Sec.\ \ref{SecVB}.
\label{figy} }
\end{figure}

Fig.\ \ref{figy} shows $\langle N(t) \rangle$ for finite systems
[Eq.\ (\ref{Ntd})]
as a function of $l/\delta$,
the scaling range of the fractal
[$\mu$ in Eq.\ (\ref{Ntd}) is an implicit function of $l$ and $\delta$, 
see Eq.\ (\ref{mu})], for different dimensions $D_F$.
The number of collisions increases of course with the scaling range 
of the fractal. For fractals small in size, say $l/\delta=100$, 
collisions with the fractals become important for dimensions $D_F$ above 
roughly 2.3.

\section{Monte Carlo simulations \label{SecVI}}

To illustrate and verify the results of the previous sections, we perform
a number of Monte Carlo simulations of random walks through fractal 
environments.

\subsection{Particle simulations: testing $p_r$ \label{SecVIA}}

In order to test the relations we found for $p_r$, we generate 
a number of fractals of different, prescribed dimensions, and 
we determine numerically the distribution of random walk 
increments.

\begin{table*}

\caption[]{For the sets $F_1$, $F_2$, $F_3$, $F_4$, the parameter $a$,
the theoretically expected  dimension $D_F$, 
the most probable nearest-neighbour-distance $\delta_\ast$,
the numerically estimated correlation dimensions ($D_{c}$),
the power-law index $\hat{\gamma}$ of $\hat{p}_r$ from the simulations, 
the analytically predicted value $\gamma$ of this index Eq.\ (\ref{pr})
(an 'e' indicates in both cases that the distribution is of exponential 
shape), 
the fraction $\hat{\nu}_{esc}$ of particles which
do not hit the fractal and escape unaffected, 
and the theoretical 
prediction $\nu_{esc}$ (Eqs.\ \ref{nuesca}, \ref{nuescb}) are listed.
\label{table1} }

\begin{tabular}{|c|c|c|r|c|r|c|c|c|c|}   \hline
 fractal &   &   &  &     &   &  &   & & \\
 set  & $n_F$ &  $a$  &  \multicolumn{1}{c|}{$D_F$}   &  $\delta_\ast$ &   
                           \multicolumn{1}{c|}{$D_{c}$}  &
                               $\hat{\gamma}$  & $\gamma$  &
                                           $\hat{\nu}_{esc}$   
                                                                & $\nu_{esc}$ \\
                         \hline\hline
$F_1$ & $100$   & $0.125$    & $1$  & $1.41$  & $1.1$    & 
                                $-2.16\pm0.07$  & $-2.0$  & $0.98$  & 
                                                                     $0.98$  \\
                           \hline
$F_2$ & $1\,000$ & $0.25$    & $1.5$ & $1.25$ & $1.6$  &
                               $-1.60\pm0.04$ & $-1.5$  & $0.95$  & 
                                                                    $0.96$  \\
                            \hline
$F_3$ & $3\,981$ & $0.31498$  & $1.8$ & $0.74$ & $1.8$   &
                               $-1.21\pm0.02$  & $-1.2$  & $0.87$ & 
                                                                     $0.82$  \\
                          \hline
$F_4$ & $10\,000$ & $0.35355$ &  $2$  & $1.06$ & $2.0$   &
                             $-1.06\pm0.02$    & $-1.0$   & $0.79$  & 
                                                                     $0.85$  \\
                          \hline
$F_5$ & $100\,000$ & $0.43528$ & $2.5$ & $0.58$ & $2.5$  & 
                                          e & e  & $0.22$  & $0.02$  \\
                                      \hline
\end{tabular}

\end{table*}

\subsubsection{Generation of test fractals \label{SecVIA1}}

The fractals we use in our simulations are generalized, 3-dimensional versions 
of the 'middle $(1-2a)$th' Cantor set (the middle 
part of length $(1-2a)$ is omitted). They are constructed with the 
method of iterated function schemes (see e.g.\ \cite{Falconer1990}), 
i.e.\ with the use of the following eight contractive maps in the
3-dimensional unit cube $[0,1] \times [0,1] \times [0,1]$
\bea
S_1(\vec x) &:=& a\vec x \nonumber \\
S_2(\vec x) &:=& a\vec x + (1-a,0,0)^T \nonumber \\
S_3(\vec x) &:=& a\vec x + (0,1-a,0)^T \nonumber \\
S_4(\vec x) &:=& a\vec x + (0,0,1-a)^T \nonumber \\
S_5(\vec x) &:=& a\vec x + (1-a,1-a,0)^T \nonumber \\
S_6(\vec x) &:=& a\vec x + (1-a,0,1-a)^T \nonumber \\
S_7(\vec x) &:=& a\vec x + (0,1-a,1-a)^T \nonumber \\
S_8(\vec x) &:=& a\vec x + (1-a,1-a,1-a)^T   ,
\label{maps}
\eea
where $0<a<0.5$ is a free parameter.
The set invariant under these contractions is a fractal 
(see e.g.\ \cite{Falconer1990}).
To generate the fractal sets in practice, a random 
point $\vec x_r$ in the unit cube is chosen and iterated with the maps
of Eq.\ (\ref{maps}), choosing at random one 
of the eight contractions at a time: the $n$-th iterate $\vec x^{(n)}$ 
is 
$\vec x^{(n)} = 
S_{i_n} ( S_{i_{n-1}} (\, ...... \, (S_{i_2} (S_{i_1} (\vec x_r)))\,.....\,))$,
with the indices $i_j$ random integer numbers between $1$ and $8$.
After a transient phase of say 1000 iterations, the iterates
$\left\{ \vec x^{(1001)},\vec x^{(1002)},\,\vec x^{(1003)},\, ....\, \right\}$
are indistinguishablely close to the underlying 
mathematical fractal, randomly distributed across it.
After their generation, the sets are shifted to have their center at the 
origin, and they are multiplied by a prescribed radial size $l$, so that 
they are contained in a sphere of radius $l$ around the origin.

Since we want to model the case of natural fractals, which show a
lower cut-off of the scaling behaviour at some scale $\delta$
(see Sec.\ \ref{SecIIA}), we must force the scaling behavior 
of the fractals we construct to break down at the scale $\delta$
--- in the way we construct the fractals, it would by chance always be 
possible that two points $\vec x^{(1000+i)}$ and $\vec x^{(1000+j)}$ are  
closer too each other than $\delta$. To achieve this, the fractals we finally 
use are defined as the subset 
$F=\left\{ \vec x^{(1000+i_1)},\vec x^{(1000+i_2)},\,\vec x^{(1000+i_3)},\, 
.....\,\vec x^{(1000+i_{n_F})} \right\}$ of all the iterates above the 
$1000$th, ($1\leq i_1 < i_2 < i_3 <\,.....,\, <i_{n_f}$), such that 
$\left\vert \vec x^{(1000+i_k)} -\vec x^{(1000+i_l)} \right\vert \geq \delta$ 
for all $k$, $l$. In practice, we just skip iterates which are closer 
than $\delta$ to at least one of the previous iterates.  
(It is to note that this forcing of a smallest inter-point distance  
is not needed in the case of natural fractals, where the  
elementary volumes they consist of cover any points which lie too close.
Stated differently, the sets we generate are {\it finite} fractals, whose 
properties we have to adjust in order them to be good models for
{\it natural} fractals; see Sec.\ \ref{SecIIA}.)

The theoretically expected dimension $D_F$ of the fractals is given as 
\begin{equation}
D_F = {\ln {1\over 8} \over \ln a}  ,
\end{equation}
for $0<a<0.5$ ($a>0.5$ implies $D_F = 3$, and the sets are not 
fractals; see e.g.\ \cite{Falconer1990}).

We generate the five sets $F_1$, $F_2$, $F_3$, $F_4$, $F_5$ listed 
in Table \ref{table1} for 
different parameters $a$ such that the corresponding 
dimensions $D_F$ are $1$, $1.5$, $1.8$, $2$, $2.5$, respectively. 
We set $\delta = 0.5$ (smallest scale) and $l=50$ (radial size), so that the 
fractal scaling behavior extends over two orders of magnitude. 
The number of points $n_F$ of the fractals should in principle 
be given by Eq.\ (\ref{nf}), but $\delta_\ast$ can be determined only 
a posteriori, after the fractals have been generated. Instead of iterating 
the generation procedure of the fractals in some way to achieve $n_F$
according to Eq.\ (\ref{nf}), we determine $n_F$ as
\begin{equation}
n_F = \left(l \over \delta \right)^{D_F}   ,
\label{nf0}
\end{equation}
since most easily and straightforwardly $\delta$, $l$ and $D_F$ can be
prescribed to the generation of the fractals.

\begin{figure}
\resizebox{\hsize}{!}{\includegraphics{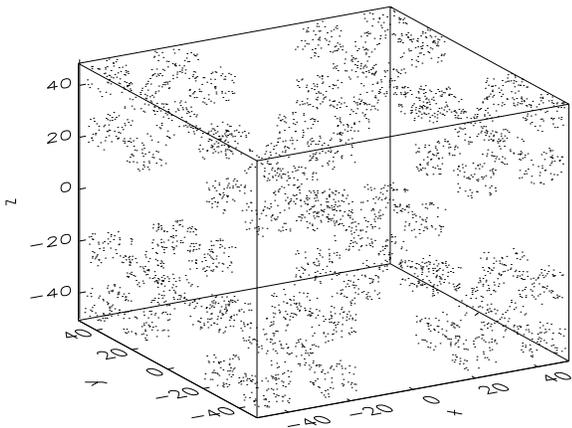}}
\caption{Projective view of the fractal set $F_3$ (dimension $D_F=1.8$; 
fine dots) we use in the 
Monte Carlo simulations (see Sec.\ \ref{SecVIA1} and Table \ref{table1} 
for its detailed properties). Over-plotted are the edges of a cube 
for better visualization.
The spatial Cartesian coordinates $x$, $y$, and $z$ are in arbitrary units.}
\label{fig1}
\end{figure}

\begin{figure}
\resizebox{\hsize}{!}{\includegraphics{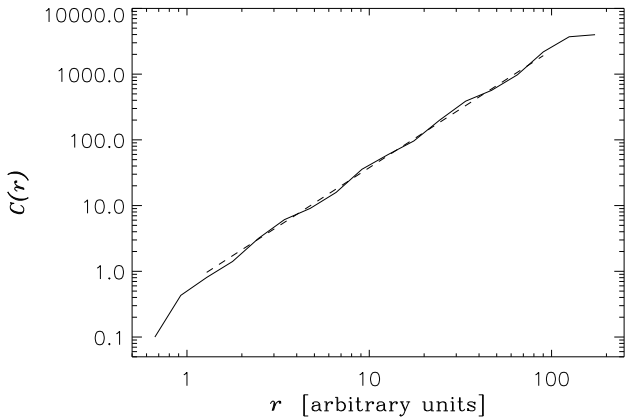}}
\caption{Correlation dimension of the set 
$F_3$ (see Table \ref{table1}): Plotted is the correlation integral $C(r)$ 
vs.\ the radius $r$ (solid), together with a power-law fit (dashed).}
\label{fig3}
\end{figure}

Fig.\ \ref{fig1} shows the set $F_3$. We confirmed the fractal dimension of the
sets by estimating their correlation dimensions (Fig.\ \ref{fig3}, 
Table \ref{table1}). Table \ref{table1} also lists the most probable 
nearest-neighbour distance $\delta_\ast$ (determined in the histograms 
of all the smallest inter-point distances, as described and illustrated in 
Sec.\ \ref{SecIIB}), which 
is needed as a parameter in the analytical relations we have derived
for $p_r$.

\begin{figure*}
\resizebox{\hsize}{!}{\includegraphics{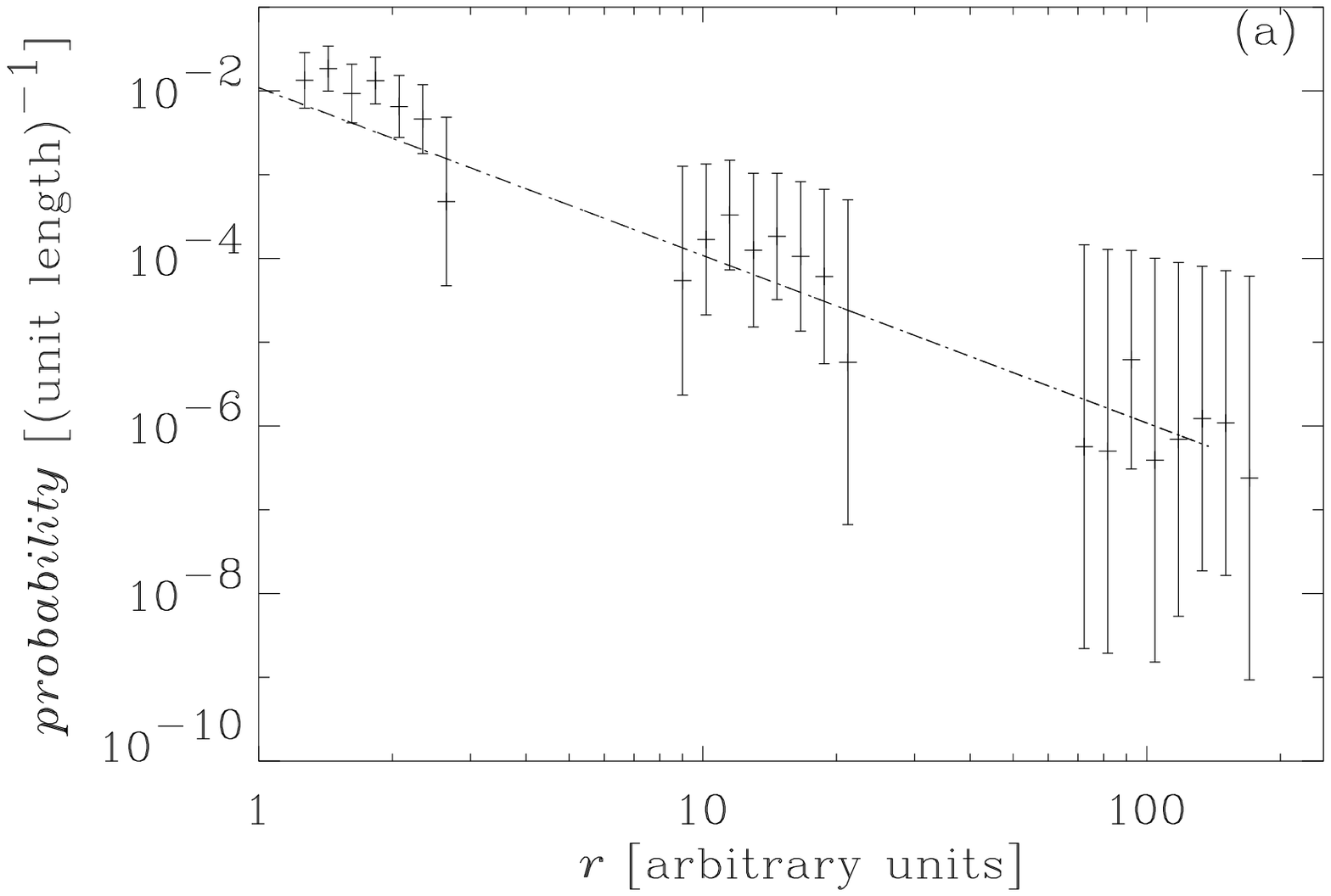}\includegraphics{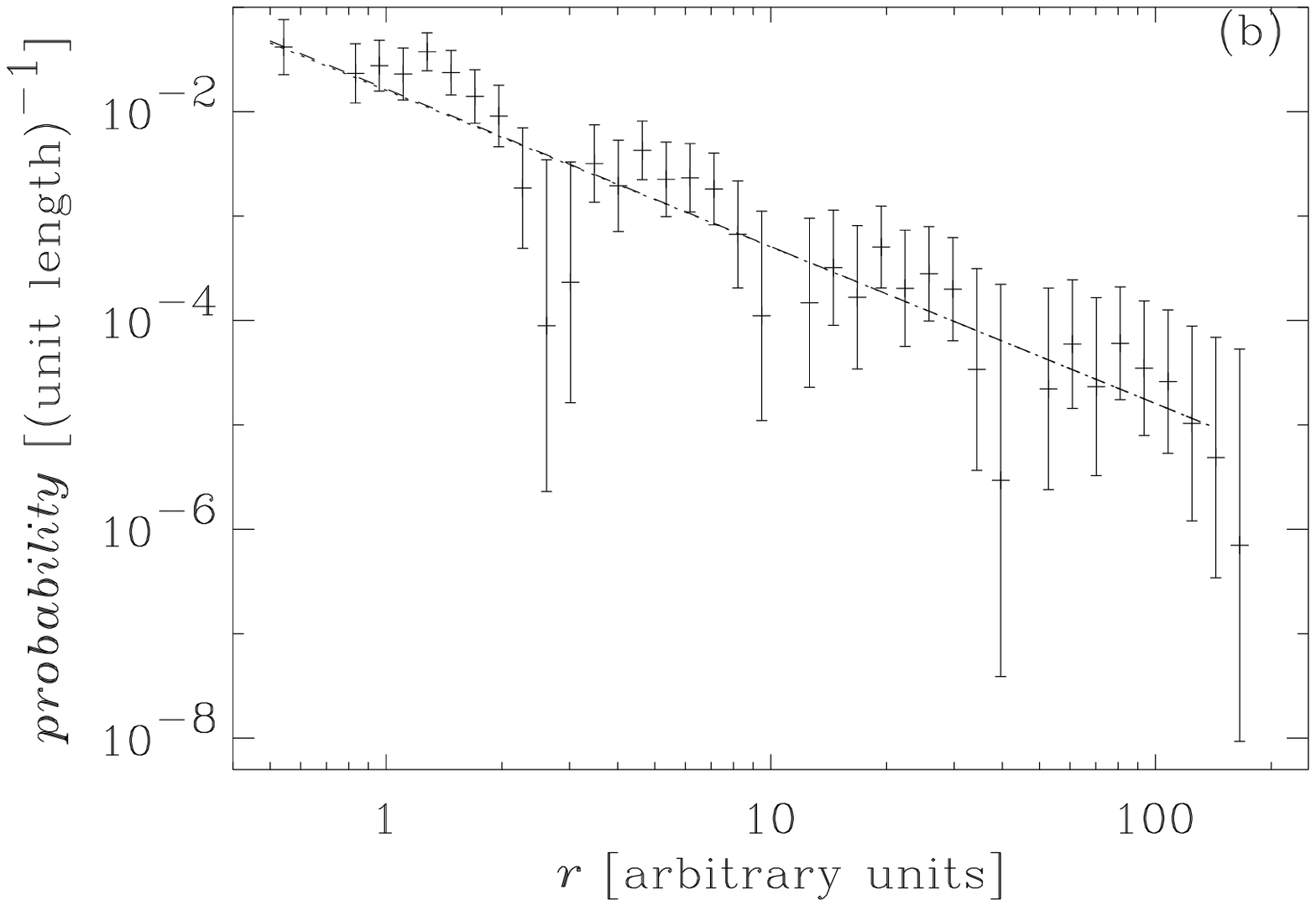}}
\resizebox{\hsize}{!}{\includegraphics{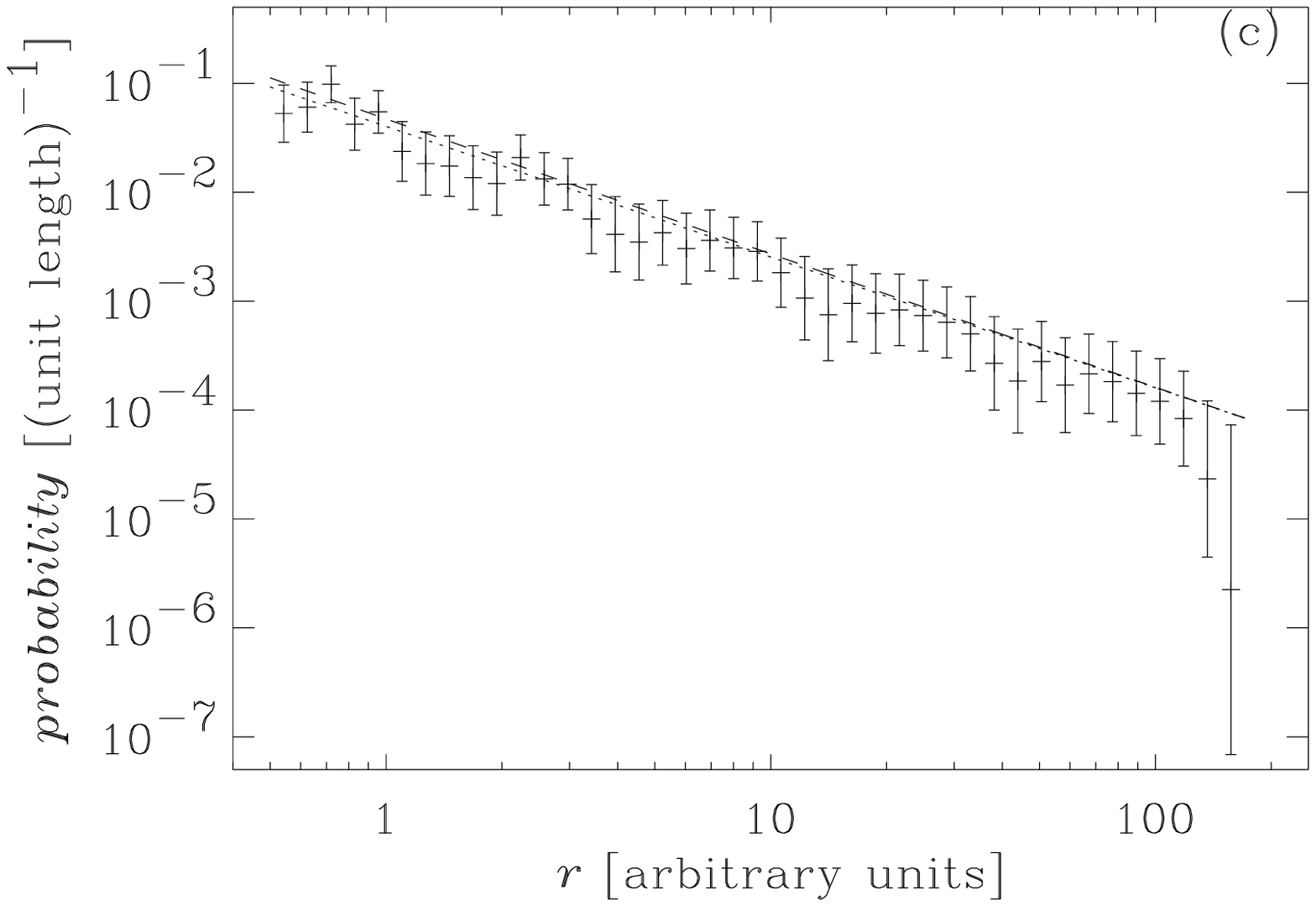}\includegraphics{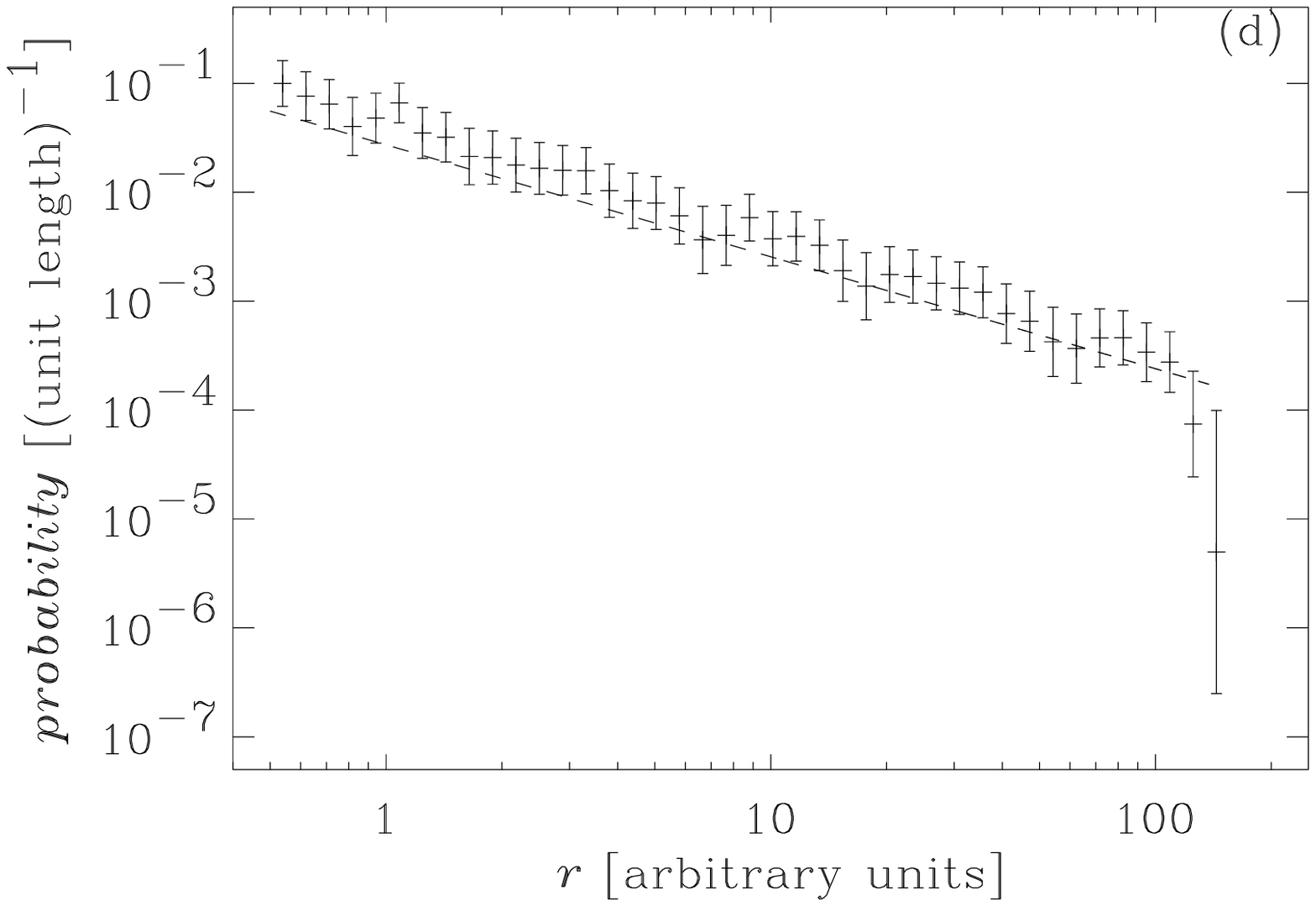}}
\resizebox{\hsize}{!}{\includegraphics{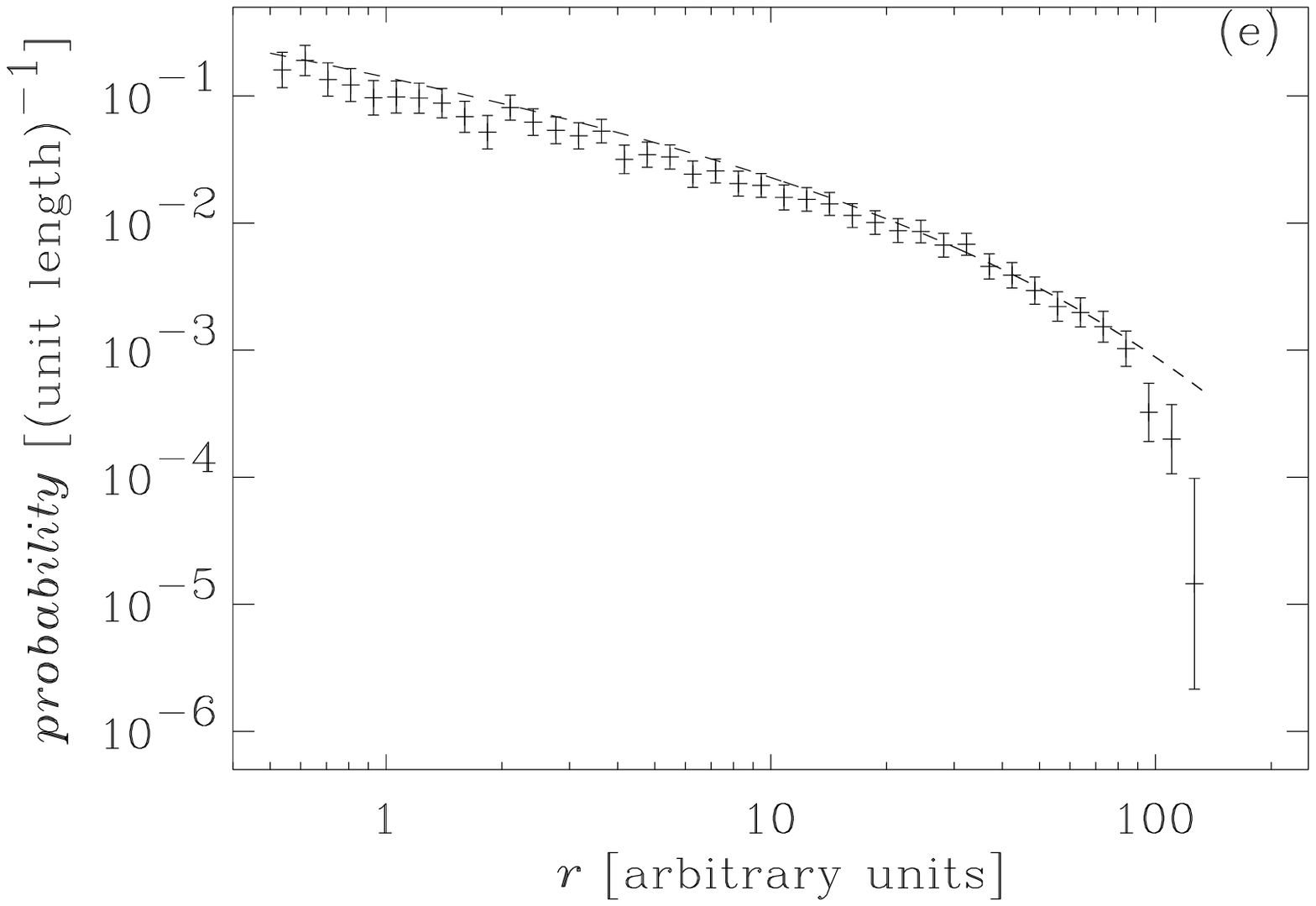}}
\caption{The probability distributions of the 
random walk increments $p_r$, as given through the 
Monte Carlo simulation ($+$ with error-bars), and as given by the 
analytical formula (Eqs.\ (\ref{pra}), (\ref{prb}); dashed), for the sets
$F_1$ (a), $F_2$ (b), $F_3$ (c), $F_4$ (d), $F_5$ (e). 
For the cases $F_1$, $F_2$, and $F_3$, also the 
approximate power-law expression for $p_r$, Eq.\ (\ref{pr}), is shown 
(dotted).}
\label{fig4}
\end{figure*}


\subsubsection{The particle simulation \label{SecVIA2}}

A number of particles $n_p$ is chosen, and for each particle we choose 
a random point $\vec x_i$ of the fractal and a random spatial
direction as initial conditions. We let each particle move into the 
random direction
and monitor at what distance it passes by another point of the fractal 
within a distance 
$\rho$, the cross-sectional radius, for the first time. 
The distances the particles travel are collected,
and their histogram $\hat{p}_r$ is constructed. 
Fig.\ \ref{fig4} shows the histograms for the sets 
$F_1$, $F_2$, $F_3$, $F_4$, $F_5$, using a cross-sectional radius 
$\rho =\delta/2 = 0.25$, together with plots of the analytically derived 
expressions for $p_r$, Eqs.\ (\ref{pra}) and (\ref{prb}), 
and of the approximate form Eqs.\ (\ref{pr}) of $p_r$ in the cases $D_F<2$. 
Table \ref{table1} lists the power-law exponents (in the case of power-laws).
The coincidence between theory and simulation is very satisfying, the 
theory describes not just the functional form correctly, but also 
the position of the simulated histograms relative to the 
$y$-axis, which means their normalization and therewith the escape rate. 
The escape rates from theory and simulations are also listed in Table 
\ref{table1}: the values are in reasonable agreement.

\begin{figure}[h]
\resizebox{\hsize}{!}{\includegraphics{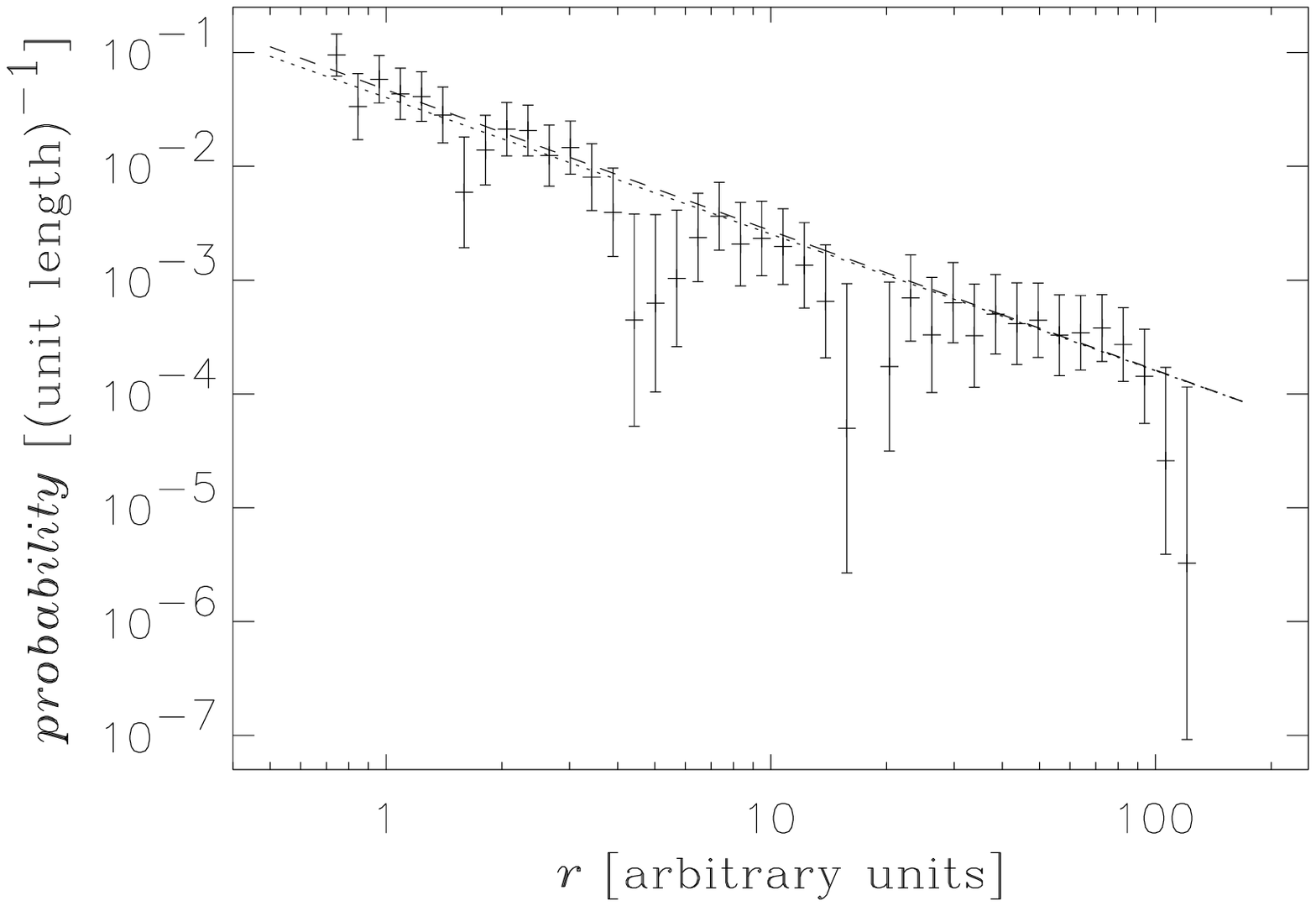}}
\caption{Same as Fig.\ \ref{fig4}, i.e.\ simulated,  
theoretical, and approximate random walk increment distributions, 
for the set $F_3$ with $D_F=1.8$. The starting points of the particles 
are though restricted to the interior of the fractal.
\label{fig5} }
\end{figure}

To investigate the influence of boundary effects, we repeated the 
simulation for the set $F_3$, with the starting points of the particles 
now restricted to the interior of the fractal.
Fig.\ {\ref{fig5}} shows the result: the boundary effects 
are obviously minor, the coincidence between simulation and theory 
is not altered.

\subsection{Particle simulations: testing the diffusive behaviour 
                                                        \label{SecVIB}}

In a second Monte Carlo simulation, we intend to confirm the 
theoretically derived results on the diffusive behaviour.
We do not use numerically generated fractals, 
since they are bound to have relatively small size, the relations 
though we want 
to verify are derived for asymptotically large systems ($l\to\infty$).
Thus, we directly use the probability distribution of flight increments 
$p_r$ [Eq.\ (\ref{pra}), (\ref{prb}), or (\ref{pr})] to determine the jump 
increments. The directions of the jumps are random.

\begin{figure}[h]
\resizebox{\hsize}{!}{\includegraphics{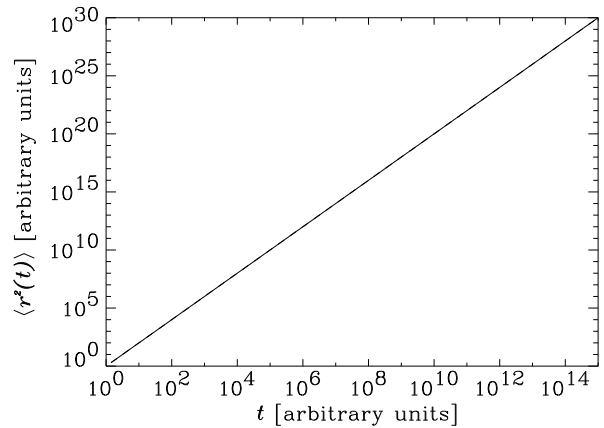}}
\caption{The mean square displacement $<r^2(t)>$ vs. time $t$ for $D_F=0.5$ 
(solid), and a power-law fit (dashed, completely coinciding with solid). 
\label{fig6} }
\end{figure}

\begin{figure}[h]
\resizebox{\hsize}{!}{\includegraphics{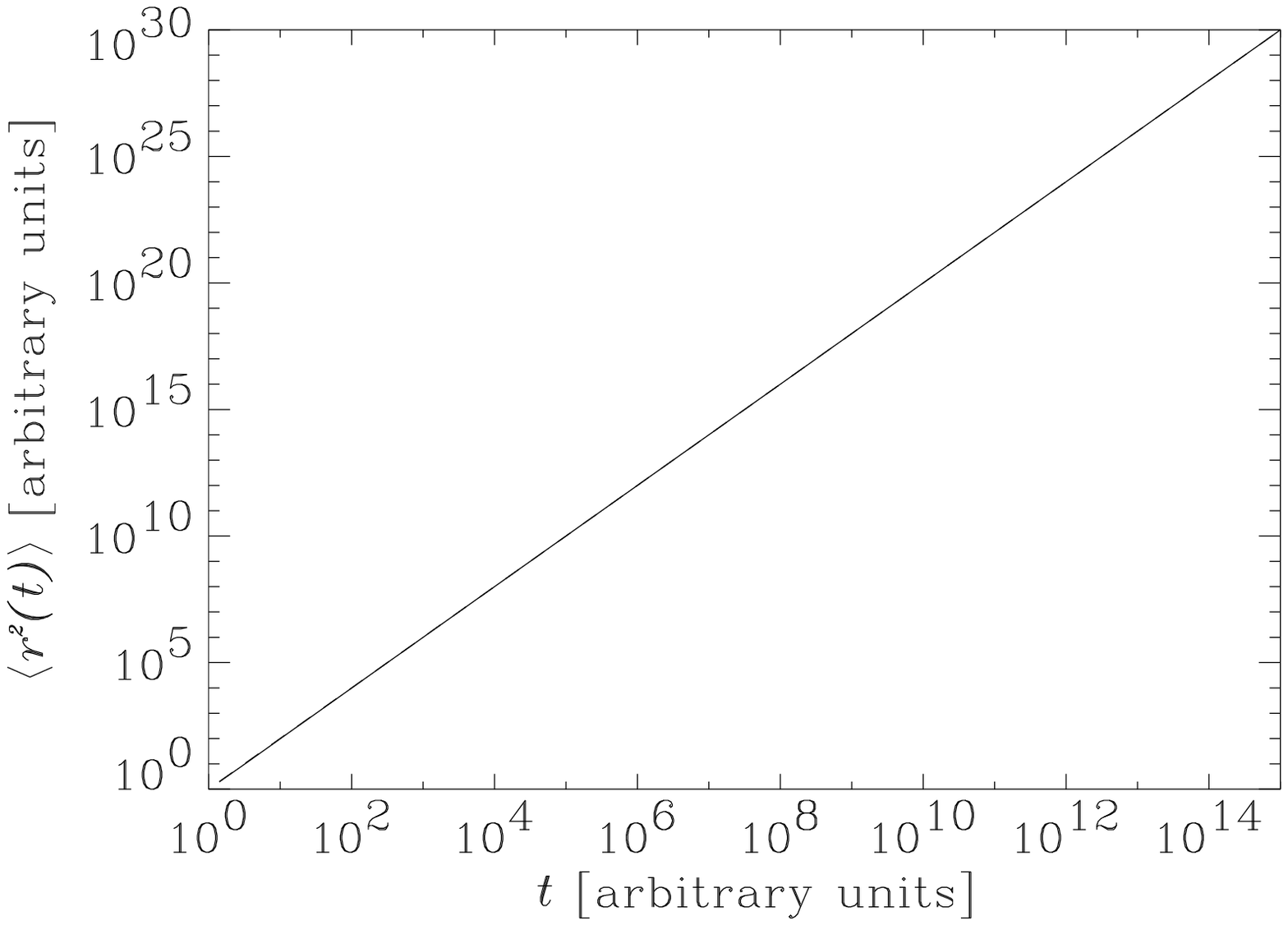}}
\caption{The mean square displacement $<r^2(t)>$ vs. time $t$ for $D_F=1.5$ 
(solid), and a power-law fit (dashed, completely coinciding with solid).
\label{fig7} }
\end{figure}

\begin{figure}[h]
\resizebox{\hsize}{!}{\includegraphics{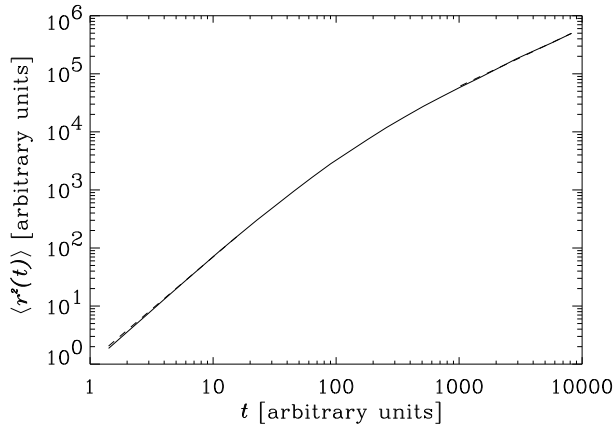}}
\caption{The mean square displacement $<r^2(t)>$ vs. time $t$ for $D_F=2.5$ 
(solid), and two power-law fits (both dashed), one in the range 
$1\leq t \leq 20$, and the other in the range $800\leq t \leq 10\,000$.
\label{fig8} }
\end{figure}

For a given dimension $D_F$, we determine first the probability 
$\nu_{esc}$ to move unaffected by the fractal forever
[Eq.\ (\ref{nuesca}) or (\ref{nuescb})]. 
All the particles start at time $t=0$ at the origin.
At the start as well as after every 'collision' with the fractal
(which in this simulation are mere turning points), 
the particles have a probability $\nu_{esc}$ 
to move for ever unaffected by the fractal on a straight line path, 
or else, with probability $1-\nu_{esc}$, they perform 
a jump of length randomly distributed according to $p_r$ 
[Eq.\ (\ref{pra}), (\ref{prb}), or (\ref{pr})] into a random direction
and 'collide' again with the fractal 
(actually they just arrive at their new turning point). 
The results are shown 
in Figs. \ref{fig6}, \ref{fig7}, and \ref{fig8} for the cases $D_F=0.5$, 
$D_F=1.5$, and $D_F=2.5$, respectively, together with power-law fits: 
The diffusion is ballistic in the cases $D_F<2$ (the index of the power-law
fits is $2$), and normal for $D_F>2$ (the index of the power-law fit at 
large times is $1$), which confirms our analytical results 
[Eqs.\ (\ref{r2t}), (\ref{r2tb})]. 

The cases $D_F=0.5$ and $D_F=1.5$ show a very unambiguous behaviour, as a 
result of the high rate for unaffected escape, which causes most particles 
not to collide anymore with the fractal
already after very few collisions, i.e.\ after relatively short time. 
For $D_F=2.5$, diffusion becomes normal only for large 
times, for small and intermediate times diffusion is enhanced:
the index of the power-law fit at small times in Fig.\ \ref{fig8} is $1.8$.

\section{Summary and Discussion \label{SecVII}}

\subsection{Summary of the results \label{SecVIIA}}

We have analytically derived the distribution of jump increments
for random walk through fractal environments, as well as 
the corresponding diffusive behaviour.
We discern between finite and asymptotically large systems, the latter
being so large that the escape rate $\nu_{esc}$ has practically settled
to its asymptotic value.
The main results are:

\medskip
\noindent {\it Fractal dimension $D_F<2$:} \\
(i)   the distribution of walk increments can be considered to be a 
      power-law with index $D_F-3$; \\
(ii)  there is always a finite rate of unaffected escape, which is usually 
      considerably large, even for asymptotically large systems; 
      the distribution 
      of jump increments is thus defective; \\
(iii) the diffusion is ballistic.

\medskip
\noindent {\it Fractal dimension $D_F>2$:} \\
(i)   the distribution of walk increments is exponentially decaying; \\
(ii)  for asymptotically  large systems, the escape rate is 
      zero, it 
      becomes positive for finite systems; \\
(iii) the diffusion is normal for large times and large systems;  \\
(iv)  even for asymptotically  large systems, there is a 
      transient phase at small
      and intermediate times where diffusion is enhanced.

\medskip\noindent
All these results have been verified with Monte-Carlo simulations.
The theory we introduced predicts in particular in a satisfying 
way the escape rate and the point where the distribution of jump increments
turns over to exponential in the cases $D_F > 2$ --- both these features 
depend very sensitively on the parameters of the model, as arguments of 
exponential functions.

The case $D_F=2$ is an exact power-law, and the escape rate is zero 
for asymptotically  large systems. We did not treat 
the diffusive behaviour of this boundary case.

\subsection{Discussion \label{SecVIIB}}

The parameters which describe the problem of random walks through 
fractal environments are the smallest distance $\delta$ between 
points of the fractal, the scale $\delta_{\ast}$ where the scaling of the 
fractal breaks down on the average, the radial size $l$ of the fractal, 
the dimension $D_F$ of the fractal, the cross-sectional radius $\rho$ of 
the points (elementary volumes)
of the fractal, and the velocity $v$ of the random walkers.
The results (jump distribution $p_r$, escape rate $\nu_{esc}$) 
do not depend on the absolute spatial scales, 
but just on the relative scales $l/\delta_\ast$ (the extent of the scaling of 
the fractal), $\delta/\delta_\ast$ (which is close to $1$, see 
Sec.\ \ref{SecIIB}), and $\rho/\delta_\ast$, 
which is in any case smaller than $1/2$ (see Sec.\ \ref{SecIIA}). 
Notably, the scaling range $l/\delta_\ast$ does not influence the
functional form of $p_r$.
The velocity $v$ is assumed to be constant and plays just a minor role
in our set-up.

The random walk in the cases $D_F<2$ is of the Levy type (Sec.\ \ref{SecIIE}). 
The distribution
of jump increments $p_r$ is though defective, i.e.\ 
not normalized to one, which implies a finite rate for unaffected escape
(Sec.\ \ref{SecIID}). Thus, even for asymptotically large systems, 
particles interact very restrictedly with fractals with dimension 
below 2, they almost do not 'see' the fractals and 
are almost not hindered on their path, in their majority they 
move unaffected on a straight line path 
already after very few collisions with the fractal.
Consequently, diffusion is  ballistic [see Eq.\ (\ref{r2t})]: 
from the beginning a considerable fraction and after some 
time the vast majority of the particles move freely according to 
$\vec r = \vec v t$, so that the square displacement from the origin becomes 
$\vec r^2 \sim t^2$. Diffusion is thus governed by the finite escape rate.

From the form of $p_r$ in the cases $D_F<2$ [Eq.\ \ref{pr}], it follows that 
$p_r$ is the steeper, the lower $D_F$ is, which implies that for the thinner 
fractals (the ones with the lower dimension) long jumps are less likely --- 
this seems paradoxical. The paradox is though resolved when taking the 
escape rate into account: the lower $D_F$ is, the more particles 
move unaffected on straight line paths for ever, so that actually long jumps 
--- including the infinite jumps along unaffected paths --- 
are more likely the lower $D_F$ is.

For dimensions $D_F$ above 3, the fractals become efficient scatterers, they 
even force normal diffusion, though only for large times.
In the regime of short and intermediate times,
diffusion is clearly different from normal, namely enhanced 
(see Fig.\ \ref{fig8}):
The distribution of jump increments $p_r$ is of power-law shape 
with an exponential turn-over [Eq.\ (\ref{pra})], and it seems 
that for intermediate times (i.e.\ small jump increments) the
power-law part of $p_r$ is essential for the diffusive behaviour, 
whereas in the large time regime the exponential roll-over starts to 
dominate.

The analytical treatment of the diffusivity we presented is valid only 
for infinitely large systems. For finite systems, $p_r$ is defective also 
in the cases $D_F>2$, there is a finite rate of unaffected escape 
(see Sec.\ \ref{SecIID}), which must be expected to modify 
the results we found here for $D_F>2$ and infinite systems,
above all in the case of relatively small systems.
For large but finite systems, our results concerning diffusion 
can be expected to remain basically valid.
The analytical study of finite
size effects on diffusion we leave for a future study,
it needs different mathematical methods than the ones applied here.

It is worthwhile noting that the distinctly different behaviour of random 
walk through fractal environments we found for the cases where $D_F$ is above
or below 2 reflects the property of {\it mathematical} fractals mentioned in 
Sec.\ \ref{SecIIA}: scattering off mathematical fractals with 
dimension below 2 is practically inexistent, with dimension above 2 it gets 
though very efficient.

The cross-sectional radius $\rho$ we used in the 
simulations was the maximal allowed value, $\rho=\delta/2$ 
(see Sec.\ \ref{SecIIA}).
Depending on the concrete application, $\rho$ might be smaller 
than $\delta/2$, which would imply that in the cases where the 
escape rate is finite, it will increase, and the behaviour of the system 
will even more be dominated by the escape rate.

The scattering process is strongly simplified in that we assume 
that the velocity is conserved in magnitude in collisions with
the fractal, we do not model the energetic aspects of the 
random walk at this stage. 

We made the assumption that there are 
no correlations between the incidence-direction and the escape-direction
for particles interacting with a point (elementary volume) of the fractal,
or more precise:
if there are correlations between the incidence and escape direction, then 
only the elementary volume should be in charge of this correlation,
it should not be caused by the over-all structure of the fractal, so that 
'seen' from the view-point of  
the fractal, incidence and escape directions appear to be random.
In plasmas though, the situation might be more complex, there may be 
a background magnetic field which guides the particles, and the electric
field residing in the scattering centers may be correlated in direction
with the magnetic field.

We assumed open boundaries, particles leave the system once 
the have reached the edge of the systems. In realistic plasma 
applications, there may well be an efficient mechanism of reinjection,
i.e.\ the particles are mirrored back into the system: In space plasmas,
magnetic mirroring at converging magnetic field topologies 
is a well known effect, and in confined plasmas with toroidal topology,
particles must be expected to reenter the fractal (turbulent) region
since they are forced to follow the closed, torus-shaped magnetic field.

Some of the histograms $\hat{p}_r$ of jump increments 
from the simulations show a more or less
strong oscillation super-imposed onto the power-law behaviour 
(see Figs.\ \ref{fig4} and \ref{fig5}), as do the estimates of the correlation 
integral (see Fig.\ \ref{fig3}). 
These oscillations are actually caused by {\it lacunarity},
i.e.\ the property of a fractal to have systematically interwoven empty 
regions (Mandelbrot in \cite{Mandelbrot1982} discusses in detail
this property of fractals). In Ref.\cite{period},  
it was shown that the scaling behaviour $n(r) \propto r^{D_F}$
for fractals (see Sec.\ \ref{SecIIB}) should actually be replaced by 
\begin{equation}
n(r) \propto r^{D_F} f(\ln r/P) ,
\end{equation}
with $f$ being an unknown periodic function of period 1. The period $P$
and the amplitude of the superimposed oscillations cannot be known a priori,
they are an inherent property of the concrete fractal under scrutiny.
We decided not to include this effect in the theory. It contains 
several parameters which are not easily estimated from a fractal, and 
in many fractals (admittedly though not in all), the amplitude of the 
oscillation is relatively small, 
the oscillation is often rather like a 'higher order correction',
and it is a reasonably good approach to 
neglect the effect --- as Fig.\ \ref{fig4} shows, our theory catches quite well
the basic features of the fractals.

\section{Conclusion \label{SecVIII}}

The theory presented here has potential applications to 
permeable media, such as plasmas (stellar atmospheres,
the magnetosphere, confined plasmas), with fractally distributed 
inhomogeneities (turbulence) which 
affect particle motion. It connects 
the respective fractal structures to random walks, and, 
eventually, to anomalous diffusion.

What we presented here is the basic analysis of random walk 
through fractal environments. A next step will be to extend the theory
by including the random walk in velocity space, 
which the particles perform in parallel to the random walk 
in direct space.
The velocity of the random walkers will no more be constant, but
it will change at the collisions with the fractal on the base
of a stochastic model for the field inhomogeneities (electric fields 
in the case of 
plasmas). This will allow to study particle acceleration in turbulent media 
through the general approach of random walks and stochastic processes.

\begin{acknowledgments}
Work performed under the Contract of Association ERB 5005 CT 99 0100 between
the European Atomic Energy Community (Euratom) and the Hellenic Rebublic.
\end{acknowledgments}

\appendix

\section{Alternative derivation of the escape rate \label{AppA}}

In this Appendix, we confirm Eqs.\ (\ref{nuesca}) and (\ref{nuescb}) 
for the escape rate $\nu_{esc}$
in an alternative way, which reveals the connection of 
$\nu_{esc}$ to the normalization $\mu$ of $p_r$ [Eq.\ (\ref{mu})]:

Since $p_r$ is the probability to travel freely a distance $r$ and then to 
collide with the fractal (see Sec.\ \ref{SecIIC}), 
$\int\limits_\delta^l p_r \, dr$ is the probability 
to hit the fractal at all for a particle which has started from a 
point of the fractal. The rate of unaffected escape is therefore alternatively
given as
\begin{equation}
\nu_{esc} = 1 - \int\limits_\delta^l p_r \, dr 
\end{equation}
so that, with the definition of $\mu$ in Eq.\ (\ref{mu}), we have 
$\nu_{esc} = 1-\mu$, and Eq.\ (\ref{numu}) follows.

The jump distribution $p_r$ can be integrated analytically:
in the case $D_F\ne 2$, the indefinite integral of $p_r$ [Eq.\ (\ref{pra})] 
is
\begin{eqnarray}
\int\limits^r p_{r\prime} dr\prime &=&
-\exp\left[   
 \frac{ D_F \rho^2 \left( \left(\frac{r}{\delta_\ast}\right)^{D_F-2} - 
       \left(\frac{\delta}{\delta_\ast}\right)^{D_F-2}\right) }
                {4(2-D_F)\delta_\ast ^2}\right] \nonumber \\
 && + \ const.   ,
\end{eqnarray}
so that we find
\begin{eqnarray}
\mu &=&
\int\limits_\delta^l p_{r\prime} dr\prime  \nonumber \\
&=& 1-\exp\left[ \frac{ D_F \rho^2   
 \left( \left(\frac{l}{\delta_\ast}\right)^{D_F-2} - 
 \left(\frac{\delta}{\delta_\ast}\right)^{D_F-2}\right) }
                     {4(2-D_F)\delta_\ast ^2} \right] .
\label{a4}
\end{eqnarray}
Eq.\ (\ref{a4}) together with Eq.\ (\ref{numu}) confirms 
Eq.\ (\ref{nuesca}). The confirmation of Eq.\ (\ref{nuescb}) is 
completely analogous.

Eq.\ (\ref{a4}) implies that $\mu\leq 1$ for any choice of the parameters 
$\delta$, $\delta_\ast$, and $l$ (with $\delta \leq \delta_\ast <l$, see
Sec.\ \ref{SecIIA}), the interpretation of $p_r$ as a probability
distribution is thus consistent.
In particular, from Eq.\ (\ref{a4}) follows $\mu < 1$ for $D_F<2$, 
and $p_r$ is always defective. For $D_F>2$, we find 
$\mu \leq 1$, where $\mu = 1$ only if $l=\infty$. 
The possibly finite escape rate ($\nu_{esc} \geq 0$) discussed in 
Sec.\ \ref{SecIID} is thus related to the fact that $\mu\leq 1$, the 
probability distribution $p_r$ is possibly defective, not necessarily 
normalized to 1.


\section{Fourier and Laplace transforming the probability distributions 
\label{AppB}}

The distributions $\psi(\vec r,t)$, $\Phi^{(c)}(\vec r,t)$, and $\Phi^{(e)}(\vec r,t)$ 
are all of the same functional form, so that 
their Fourier-Laplace transforms are analogous.
We demonstrate the way we calculate 
these Fourier-Laplace transforms on the example of the general function 
$\chi(\vec r,t)$, which is of the form 
\begin{equation}
\chi(\vec r,t) = \chi(r) \delta(t-r/v) ,
\label{k0}
\end{equation}
as are $\psi(\vec r,t)$, $\Phi^{(c)}(\vec r,t)$, and $\Phi^{(e)}(\vec r,t)$, 
with $\delta\leq r\leq \infty$, $\delta/v \leq t \leq \infty$, and where 
$r:=\vert \vec r\vert $.
Also the distribution $\Phi^{(0)}(\vec r,t)$ is of the form Eq.\ (\ref{k0}),
and basically the expressions we derive for $\chi(\vec r,t)$ are also valid for
$\Phi^{(0)}(\vec r,t)$, with some modifications though, 
since $\Phi^{(0)}(\vec r,t)$ has a finite support ($0\leq r\leq \delta$).
The treatment of $\Phi^{(0)}(\vec r,t)$ 
will be presented in App.\ \ref{AppB4}.

\subsection{The Fourier transforms $\chi(\vec r,t)$ \label{AppB1}}

The Fourier transform of $\chi(\vec r,t)$ in spherical 
coordinates $(r,\theta,\phi)$ is defined as 
\bea
\chi(\vec k,t) 
&=& 
\int\limits \!\! d^3r \,
\chi(\vec r,t)  \, e^{i\vec k\cdot \vec r}   \\
&=& 
\int\limits \!\! r^2 \sin\theta\, d\phi\, d\theta\, dr \,
\chi(r) \delta(t-r/v)  \, e^{i\vec k\cdot \vec r} 
\label{k1} \\
&=& 
\int\limits_\delta^\infty \!\! r^2 dr\,    
\chi(r) \delta(t-r/v)   
\nonumber 
\\
& & \qquad\qquad
         \times \int\limits_0^\pi \!\! d\theta\,\sin\theta\, e^{i\vec k\cdot \vec r} 
\!\! \int\limits_{0}^{2\pi} \!\! d\phi  ,
\label{k2}
\eea
where we have explicitly introduced the lower limit $\delta$ for the
$r$-integral, below which $\chi(\vec r,t)$ is zero.
For the $\theta$-integral, we can assume without loss of generality that
$\vec k \vert\vert \hat z$, so that $\vec k \cdot \vec r = kr\cos\theta$, where 
$k:=\vert \vec k \vert$. 
Substituting furthermore $x:= \cos\theta$, the $\theta$-integral becomes
\begin{equation}
\int\limits_0^\pi \!\! d\theta\, \sin\theta\,  e^{i k r\cos\theta}
= \int\limits_{-1}^1 \!\! dx\, e^{i k r x}
= {2\over kr} \sin (kr) ,
\label{k3}
\end{equation}
so that 
\beq
\chi(\vec k,t) 
=
4\pi \int\limits_\delta^\infty \!\! r dr\,    
\chi(r) \delta(t-r/v) \frac{\sin k r}{k}  .
\label{k4}
\eeq
The $\delta$-function in Eq.\ (\ref{k4}) implies firstly $r=vt$, secondly 
$t\geq\delta/v$ (since $r\geq \delta$), and thirdly that the entire expression 
must be 
multiplied by $v$ (as a substitution 
$r\to \zeta := t-r/v$ would bring forth), so that $\chi(\vec k,t)$ becomes
\beq
\chi(\vec k,t) = 4\pi v^2 \,t\, \chi(vt)\, \frac{\sin kvt}{k}  ,
\label{k5}
\eeq
with $t\geq \delta/v$.
Assuming $kvt<<1$ (see Sec.\ \ref{SecIIIB}), we approximate 
$\sin kvt \approx kvt - {1\over 6} (kvt)^3$, which yields
\beq
\chi(\vec k,t) \approx 4\pi v^3\, t^2\, \chi(vt) 
                            - \frac{4\pi}{6} k^2 v^5\, t^4\, \chi(vt)  .
\label{k6}
\eeq

For conciseness, it is useful to introduce the marginal probability 
distribution $\lambda(t)$ of $\chi(\vec r,t)$, integrated over space,
\bea
\lambda(t) &:=& \int \chi(\vec r,t) \, d^3r 
\label{k61} \\
&=& \int \chi(r) \delta(t-r/v) \, r^2 \sin\theta \, dr \, d\phi \, d\theta
\label{k62}   \\
&=& 4\pi \int \chi(r) \delta(t-r/v) \, r^2 \, dr  ,
\label{k63}
\eea
where in Eq.\ (\ref{k62}) we used spherical coordinates, and in Eq.\ (\ref{k63})
we exploited the spherical symmetry. 
The $r$-integration of the $\delta$-function implies $r=vt$ and an over-all 
multiplication by $v$, so that finally
\beq
\lambda(t)  = 4\pi v^3 t^2 \chi(vt) ,
\label{k71}
\eeq
with $t\geq \delta/v$.
With the aid of $\lambda(t)$, $\chi(\vec k,t)$ [Eq.\ (\ref{k5})] can now be written
as
\beq
\chi(\vec k,t) = v^{-1} t^{-1} \lambda(t) \frac{\sin kvt}{k} ,
\label{k81}
\eeq
and the approximate form [Eq.\ (\ref{k6})] writes 
\beq
\chi(\vec k,t) \approx \lambda(t) - \frac{1}{6} k^2 v^2t^2\lambda(t) .
\label{k8}
\eeq

\subsection{The Laplace transform of $\chi(\vec k,t)$ \label{AppB2}}

Through Eq.\ (\ref{k8}), the Laplace transform of $\chi(\vec k,t)$,
defined as
\beq
\chi(\vec k,s) = \int\limits_0^\infty dt\, \chi(\vec k,t) e^{-st} ,
\label{k9}
\eeq
reduces for small $k$ to the Laplace 
transforms of $\lambda(t)$ and $t^2\lambda(t)$,
\beq
\chi(\vec k,s) \approx   \int\limits_{\delta/v}^\infty dt\, \lambda(t)\,e^{-st} 
        - \frac{1}{6} k^2 v^2 \int\limits_{\delta/v}^\infty dt\, t^2 \lambda(t)\, 
                                                             e^{-st}   .
\label{k10}
\eeq

\subsubsection{The Laplace transform of $\lambda(t)$ \label{AppB2a}}

Assuming $s<<1$, we approximate the Laplace transform $\lambda(s)$ of $\lambda(t)$,
\beq
\lambda(s)  = \int\limits_{\delta/v}^{\infty} \lambda(t) e^{-st} \, dt   ,
\label{k11}
\eeq
by expanding $\lambda(s)$ around $s=0$ according to 
\begin{equation}
\lambda(s) \approx
\lambda(s)\vert_{s=0} + s\cdot \frac{d}{ds} \lambda(s)\vert_{s=0}   ,
\label{lambdaexpand}
\end{equation}
so that from Eq.\ (\ref{k11})
\bea
\lambda(s) &\approx&
\int\limits_{\delta/v}^{\infty} \lambda(t) \, e^{-st} \, dt \Bigg\vert_{s\to0}
                                                          \nonumber \\
           &&       -  s \cdot \int\limits_{\delta/v}^{\infty} t\,\lambda(t)\,e^{-st} 
                                        \, dt  \Bigg\vert_{s\to0}
\label{k12}    \\
&=& B^{(0)}(s)\Big\vert_{s\to 0} - s\cdot B^{(1)}(s)\Big\vert_{s\to 0} ,
\label{k122}
\eea
where for convenience we have introduced the functions
\beq
B^{(n)}(s) := \int\limits_{\delta/v}^{\infty} 
                                        t^n \lambda(t)e^{-st} \, dt ,
\label{Bns}
\eeq
with the integer parameter $n=0,1,2,3,...$.

\subsubsection{The Laplace transform of $t^2 \lambda(t)$ \label{AppB2b}}

Analogously to the case of $\lambda(t)$, we determine the Laplace transform 
of $t^2 \lambda(t)$,
\beq
L[t^2 \lambda(t)](s) = \int\limits_{\delta/v}^\infty t^2 \lambda(t) \,e^{-st}\,dt ,
\label{k13}
\eeq
by approximating in the way of Eq.\ (\ref{lambdaexpand}),
\bea
L[t^2 \lambda(t)](s) &\approx&
\int\limits_{\delta/v}^{\infty} t^2\, \lambda(t)\, e^{-st}\, dt \Bigg\vert_{s\to 0} 
                                               \nonumber \\
              &&    -  s \cdot \int\limits_{\delta/v}^{\infty} t^3 \,\lambda(t) 
                                            \, e^{-st}\, dt \Bigg\vert_{s\to 0}
\label{k14} \\
&=& B^{(2)}(s)\Big\vert_{s\to 0} - s\cdot B^{(3)}(s)\Big\vert_{s\to 0}   ,
\label{k144}
\eea
where we have again identified the functions $B^{(n)}(s)$ [see Eq.\ (\ref{Bns})].

Inserting Eqs.\ (\ref{k122}) and (\ref{k144}) into Eq.\ (\ref{k10}) yields
for $\chi(\vec k,s)$ 
\bea
\chi(\vec k,s) &=& B^{(0)}(s)\Big\vert_{s\to 0} - s\cdot B^{(1)}(s)\Big\vert_{s\to 0}
                                                  \nonumber \\
-  k^2 & \frac{v^2}{6} &
 \left(B^{(2)}(s)\Big\vert_{s\to 0} - s\cdot B^{(3)}(s)\Big\vert_{s\to 0}\right)   .
\label{chichi}
\eea
The problem of Laplace transforming $\chi(\vec k,t)$ is thus reduced to 
evaluating the functions $B^{(n)}(s)$ for $s\to 0$ and $n=0,1,2,3$.

\subsection{Evaluating the functions $B^{(n)}(s)$ for $s\to 0$ \label{AppB3}}

The function $B^{(n)}(s)$ at $s=0$,
\beq
B^{(n)}(s)\Big\vert_{s= 0} = \int\limits_{\delta/v}^\infty t^n \lambda(t)\,dt
=:\langle T^n\rangle_\lambda ,
\eeq
is the $n$th moment $\langle T^n\rangle_\lambda$ of $\lambda(t)$.
In particular, $B^{(0)}(s)\Big\vert_{s= 0}$ is the 
normalization $\mu_\lambda$ of $\lambda(t)$,
and we note that 
\begin{eqnarray}
B^{(0)}(s)\Big\vert_{s=0} &\equiv&
\int\limits_{\delta/v}^\infty \lambda(t) \,dt \\
&=& \int \chi(\vec r,t) \,d^3r \, dt,  \label{kaa1} \\
&=& \int\limits_{\delta}^\infty \chi(\vec r) \delta(t-r/v) \,d^3r\,dt  \label{kaa2}  \\
&=& \int\limits_{\delta}^\infty \chi(\vec r) \,d^3r \label{kaa3}  \\
&=& \int\limits \chi_r \,dr = \mu_\lambda   , \label{chinorm}
\end{eqnarray} 
where in Eq.\ (\ref{kaa1}) we basically repeated the definition of 
$\lambda(t)$ [Eq.\ (\ref{k61})], in Eq.\ (\ref{kaa2}) we inserted the generic 
form of $\chi(\vec r,t)$ [Eq.\ (\ref{k0})], in Eq.\ (\ref{kaa3}) we did
the $\tau$-integration, and in Eq.\ (\ref{chinorm})
we introduced $\chi_r$, the marginal spatial probability
distribution of $\chi(\vec r)$, integrated over solid angle:
$\chi_r := \int \chi(\vec r)\,d\sigma$ (in analogy 
to how $p_r$ is related to $p(\vec r)$, see Sec.\ \ref{SecIII}). 
The normalizations of $\lambda(t)$, 
$\chi(\vec r,t)$, $\chi(\vec r)$, and $\chi_r$ are thus identical and are 
represented by $\mu_\lambda$.
In the case where $\chi(\vec r, \tau)$  represents $\psi(\vec r,t)$, 
$\lambda(t)$ corresponds to $\varphi(t)$, and $\mu_\lambda$ is called $\mu$, see 
Sec.\ \ref{SecIID}.

If all the moments $\langle T^n\rangle_\lambda$ are finite up to $n=3$, 
Eq.\ (\ref{k122}) can be written
\begin{equation}
\lambda(s) \approx \mu_\lambda -  s \cdot \langle T \rangle_\lambda ,
\label{k16}
\end{equation}
and if $\lambda(t)$ is normalized to one, then we have furthermore 
$\mu_\lambda=1$.
(The first moment $\langle T \rangle_\lambda$ of $\lambda(t)$
in the case where $\chi(\vec r,t)$ represents $\psi(\vec r,t)$ corresponds to 
the expected time spent in a single jump increment.) With finite second and 
third moments, Eq.\ (\ref{k144}) becomes
\begin{equation}
L[t^2 \lambda(t)](s) \approx \langle T^2 \rangle_\lambda - 
                                  s \cdot \langle T^3 \rangle_\lambda .
\label{k17}
\end{equation}

Eqs.\ (\ref{k16}) and (\ref{k17}) are formal in the sense that the 
moments $\mu_\lambda$, $\langle T \rangle_\lambda$, 
$\langle T^2 \rangle_\lambda$, 
and $\langle T^3 \rangle_\lambda$ do not necessarily exist, they may be 
infinite. 
To determine the expressions $B^{(n)}(s)$ for $s\to 0$ and the moments of 
$\lambda(t)$, if they exist, we have to specify the different cases which 
$\chi(\vec r,t)$ and $\lambda(t)$ represent.

\subsubsection{The case $D_F>2$ \label{AppB3a}}

For $D_F>2$, $\chi(\vec r,t)$ represents
$\psi(\vec r,t)$ or $\Phi^{(c)}(\vec r,t)$. 
Using the relation Eq.\ (\ref{k71}), we find from Eq.\ (\ref{psiDgt2})
in the case of $\psi(\vec r,t)$ that 
\begin{equation}
\lambda^{(\psi)}(t) = C v^{D_F-2}\exp\left[-\beta (vt)^{D_F-2}\right] 
                                          t^{D_F-3} ,
\label{la1}
\end{equation}
and in the case of $\Phi^{(c)}(\vec r,t)$ from Eq.\ (\ref{PhicDgt2}) that 
\beq
\lambda^{(\Phi^{(c)})}(t) = 
\frac{Cv}{\beta(D-2)}  \exp\left[-\beta (vt)^{D_F-2}\right]  .
\eeq

In both cases, $\lambda(t)$ is of the form 
$\lambda(t) \sim \exp\left[-\beta (vt)^{D_F-2}\right] \cdot t^\alpha$, 
with $\alpha$ a corresponding constant,
so that the expressions
$B^{(n)}(s)\Big\vert_{s\to 0}$ [see Eq.\ (\ref{Bns})] turn to integrals 
of the form 
\beq
B^{(n)}(s) \Big\vert_{s\to 0}
\sim  \int\limits_{\delta/v}^\infty t^{n+\alpha} 
\exp\left[-\beta (vt)^{D_F-2}\right] e^{-st}\, dt \Bigg\vert_{s\to 0}  .
\label{lala1}
\eeq
The exponential guarantees that 
the integrals are finite, for $s\to 0$ and $n=0,1,2,3$, Eqs.\ (\ref{k16}) and 
(\ref{k17}) are thus valid, and $\chi(\vec k,s)$ is determined through 
Eq.\ (\ref{chichi}).

\subsubsection{The case $D_F<2$ \label{AppB3b}}

For $D_F<2$, the moments of $\lambda(t)$ can be infinite.
$\chi(\vec r,t)$ represents 
$\psi(\vec r,t)$, $\Phi^{(c)}(\vec r,t)$, and $\Phi^{(e)}(\vec r,t)$.
Through Eq.\ (\ref{k71}), the corresponding functions $\lambda(t)$ are
given for $\psi(\vec r,t)$ from Eq.\ (\ref{psirtC}) as 
\beq
\lambda^{(\psi)}(t) = C v^{D_F-2} t^{D_F-3} ,
\label{la2}
\eeq
for $\Phi^{(c)}(\vec r,t)$ from Eq.\ (\ref{ph1}) as 
\beq
\lambda^{(\Phi^{(c)})}(t) = \frac{Cv^{D_F-1}}{2-D_F} t^{D_F-2} ,
\eeq
and for $\Phi^{(e)}(\vec r,t)$ from Eq.\ (\ref{Phirt}) as 
\beq
\lambda^{(\Phi^{(e)})}(t) = \nu_{esc} v .
\eeq 
In all cases, $\lambda(t)$ is of a pure power-law form, $\lambda(t)\sim t^\alpha$, 
and the expressions
$B^{(n)}(s)\Big\vert_{s\to 0}$ [$n=0,1,2,3$; see Eq.\ (\ref{Bns})] turn to integrals 
of the form 
\beq
B^{(n)}(s) \Big\vert_{s\to 0} 
\sim  \int\limits_{\delta/v}^\infty t^{n+\alpha} e^{-st}\, dt \Bigg\vert_{s\to 0}  .
\eeq
If $n+\alpha < -1$, then the integrals are finite for $s=0$ and just equal
the $n$th moment, 
\beq
B^{(n)}(s) \Big\vert_{s= 0} 
\sim  \int\limits_{\delta/v}^\infty t^{n+\alpha} \, dt \sim \langle T^n \rangle_\lambda  .
\label{Bnsa}
\eeq

For $n+\alpha \geq -1$, $B^{(n)}(s)\Big\vert_{s\to 0}$ is infinite,
and we determine the exact divergence behaviour 
by the substitution $t \to y:= st$,
\bea
B^{(n)}(s)\Big\vert_{s\to 0} &\sim&
\int\limits_{\delta/v}^\infty t^{n+\alpha} e^{-st} \, dt \Bigg\vert_{s\to 0} \\
&=& \int\limits_{s\delta/v}^\infty \left(\frac{y}{s}\right)^{n+\alpha} 
                                           e^{-y} \, \frac{dy}{s} 
                                                \Bigg\vert_{s\to 0} \\
&=& \frac{1}{s^{n+\alpha+1}} \Big\vert_{s\to 0}
      \int\limits_{s\delta/v}^\infty y^{n+\alpha} e^{-y} \, dy \Bigg\vert_{s\to 0}  .
\label{k18}
\eea
The integral in Eq.\ (\ref{k18}) is finite and approaches $\Gamma(n+\alpha+1)$ for 
$s\to 0$ as long as $n+\alpha > -1$, 
where $\Gamma(.)$ is Euler's Gamma-function, so that 
\beq
B^{(n)}(s)\Big\vert_{s\to 0} \sim
\frac{1}{s^{n+\alpha+1}} \Gamma(n+\alpha+1)   ,
\label{Bnsb}
\eeq
for $n+\alpha > -1$.

Eqs.\ (\ref{Bnsa}) and (\ref{Bnsb}) determine $\chi(\vec k,s)$ through 
Eq.\ (\ref{chichi}).

\subsection{The Fourier Laplace transform of $\Phi^{(0)}(\vec r,t)$ \label{AppB4}}

The distribution $\Phi^{(0)}(\vec r,t)$ has the same functional form as 
$\chi(\vec r,t)$ [Eq.\ (\ref{k0})], just that its support is finite.
It can thus be treated analogous to $\chi(\vec r,t)$, and 
its Fourier Laplace transform is given by Eq.\ (\ref{chichi})
on replacing the functions $B^{(n)}(s)$ by the functions $\bar{B}^{(n)}(s)$,
\beq
\bar{B}^{(n)}(s) = \int\limits_0^{\delta/v} t^n\,\lambda(t)\,e^{-st}\,dt .
\eeq

The marginal probability distribution $\lambda^{(\Phi^{(0)})}(t)$ is given 
through Eqs.\ (\ref{k71}) and (\ref{Phi00}) (note that $\Phi^{(0)}(\vec r,t)$ is 
the same for $D_F>2$ and $D_F<2$),
\beq
\lambda^{(\Phi^{(0)})}(t) = v  ,
\eeq
with $0 \leq t \leq \delta/v$ (from $0 \leq r \leq \delta$).
The expressions
$\bar{B}^{(n)}(s)\Big\vert_{s\to 0}$ ($n=0,1,2,3$) to be determined take the 
form
\beq
B^{(n)}(s) \Big\vert_{s\to 0} 
\sim  \int\limits_0^{\delta/v} t^{n}  e^{-st}\, dt \Bigg\vert_{s\to 0} ,
\eeq
which are obviously finite for $n=0,1,2,3$, the cases needed to 
determine $\Phi^{(0)}(\vec k,s)$ through Eq.\ (\ref{chichi}).

\subsection{The Laplace transform of $\varphi(\tau)$ \label{AppB5}}

The Laplace transform of $\varphi(\tau)$ (Secs.\ \ref{SecIVB} and \ref{SecVB})
is given by Eq.\ (\ref{k122}). In the case $D_F>2$, the function
$\lambda(t)$ is given by Eq.\ (\ref{la1}), with the expressions
$B^{(n)}(s)$ for $s\to 0$ evaluated according to Eq.\ (\ref{lala1}).
In the case $D_F<2$, $\lambda(t)$ is given by Eq.\ (\ref{la2}), and
again Eq.\ (\ref{k122}) yields the Laplace transform of $\varphi(t)$, 
by using Eqs.\ (\ref{Bnsa}) or (\ref{Bnsb}) to determine $B^{(n)}(s)$ for $s\to 0$.






\newpage 


\begin{thebibliography}{}


\bibitem{Montroll1965}
E.W.\ Montroll, G.H.\ Weiss, J.\ Math.\ Phys.\ 6, 167 (1965).

\bibitem{BTW}
P.\ Bak, C.\ Tang, K.\ Wiesenfeld, Phys.\ Rev.\ Lett.\ 59, 381 (1987);
P.\ Bak, C.\ Tang, K.\ Wiesenfeld, Phys.\ Rev.\ A 38, 364 (1988).

\bibitem{Isliker2001}
H.\ Isliker, L.\ Vlahos, (2001) (unpublished result);
S.W.\ McIntosh, P.\ Charbonneau, T.J.\ Bogdan, H.-L.\ Liu, J.P.\ Norman,
Phys.\ Rev.\ E 65, 6125 (2002).

\bibitem{Isliker}
H.\ Isliker, A.\ Anastasiadis, L.\ Vlahos, Astron.\ and Astrophys.\ 363, 1134 
(2000);
H.\ Isliker, A.\ Anastasiadis, L.\ Vlahos, Astron.\ and Astrophys.\ 377, 1068 
(2001).

\bibitem{Lu}
T.E.\ Lu, R.J.\ Hamilton, Ap.\ J.\ 380, L89 (1991);
T.E.\ Lu, R.J.\ Hamilton, J.M.\ McTiernan, K.R.\ Bromund, Ap.\ J.\ 412, 
841 (1993);
L.\ Vlahos, M.\ Georgoulis, R.\ Kluiving, P.\ Paschos,
Astron.\ and Astrophys.\ 299, 897 (1995).

\bibitem{magentosphere}
S.\ Chapman, N.\ Watkins, Space Science Reviews 95, 293 (2001).

\bibitem{Drysdale1998}
P.M.\ Drysdale, P.A.\ Robinson, Phys.\ Rev.\ E 58, 5382 (1998).

\bibitem{Shlesinger1987}
M.F.\ Shlesinger, B.J.\ West, J.\ Klafter, Phys. Rev. Lett.\ 58, 1100 (1987).

\bibitem{SOC}
P.A.\ Politzer, Phys.\ Rev.\ Lett.\ 84, 1192 (2000);
S.C.\ Chapman, R.O.\ Dendy, B.\ Hnat, Phys.\ Rev.\ Lett.\ 86, 2814 (2001);
B.A.\ Carreras, D.\ Newman, V.E.\ Lynch, P.H.\ Diamond,  
Phys.\ of Plasmas 3, 2903 (1996).

\bibitem{Carbone}
E.\ Spada et al., Phys.\ Rev.\ Lett.\ 86, 3032 (2001);
V.\ Antoni et al., Phys.\ Rev.\ Lett.\ 87, 5001 (2001).

\bibitem{confined}
R.\ Balescu, Phys.\ Rev. E 51, 4807 (1995);
E.\ Barkai, J.\ Klafter, in \textit{Lecture Notes in Phys.\ 511}, edited by 
S.\ Benkadda, G.M.\ Zaslavsky, (Springer-Verlag, Berlin, 1998), p.\ 373;
G.\ Zimbardo, A.\ Greco, P.\ Veltri, Phys.\ of Plasmas 7, 1071 (2000).

\bibitem{Bak2001}
P.\ Bak, K.\ Chen, Phys.\ Rev.\ Lett.\ 86, 4215 (2001);

\bibitem{frac}
B.\ O'Shaughnessy, I.\ Procaccia, Phys.\ Rev.\ Lett.\ 54, 455 (1985);
B.\ O'Shaughnessy, I.\ Procaccia, Phys.\ Rev.\ A 32, 3073 (1985);
A.\ Blumen, J.\ Klafter, G.\ Zumofen, Phys.\ Rev.\ B 28, 6112 (1983);
L.\ Acedo, S.B.\ Yuste, Phys.\ Rev.\ E 57, 5160 (1998).

\bibitem{sand}
B.A.\ Carreras, V.E.\ Lynch, D.E.\ Newman, G.M.\ Zaslavsky, 
Phys.\ Rev.\ E 60, 4770 (1999);
P.\ B\'antay, I.M.\ J\'anosi, Phys.\ Rev.\ Lett.\ 68, 2058 (1992);
M.\ Bogu\~n\'a, \'A.\ Corral, Phys.\ Rev.\ Lett.\ 78, 4950 (1997).

\bibitem{Mandelbrot1982}
B.B.\ Mandelbrot, \textit{The Fractal Geometry of Nature} (Freeman, New York, 1982).

\bibitem{Sinai1980}
Ya.G.\ Sinai, in \textit{Nonlinear Dynamics}, edited by R.H.G.\ Helleman,
Annals of the New York Academy of Sciences vol.\ 357 (New York, 1980), 
p.\ 143.

\bibitem{Zumofen1993}
G.\ Zumofen, J.\ Klafter, Phys.\ Rev.\ E 47, 851 (1993).

\bibitem{math}
J.\ Klafter, A.\ Blumen, M.F.\ Shlesinger, Phys.\ Rev.\ A 35, 3081 (1987);
A.\ Blumen, G.\ Zumofen, J.\ Klafter, , Phys.\ Rev.\ A 40, 3964 (1989).

\bibitem{Feller1971}
W.\ Feller, \textit{An Introduction to Probability Theory and its Applications},
Vol.\ 2, 2nd edition (Wiley, New York, 1971).

\bibitem{Falconer1990}
K.\ Falconer, \textit{Fractal Geometry} (John Wiley \& Sons, Chichester, 1990).

\bibitem{period}
R.\ Badii, A.\ Politi, Phys.\ Lett.\ A 104, 303 (1984); 
L.A.\ Smith et al.\ , Phys.\ Lett.\ A 114, 465 (1986);
H.\ Isliker, Phys.\ Lett. A 169, 313 (1992).

\bibitem{Hughes1995}
Hughes, B.D., \textit{Random Walks and Random Environments, vol.\ 1: Random Walks} 
(Clarendon Press, Oxford, 1995)

\bibitem{Morse}
Morse, P.M., Feshbach, H. \textit{Methods of Theoretical Physics, part I} 
(McGraw-Hill, New York, 1953)





\end{thebibliography}
\end{document}